\documentclass[a4paper,11pt]{article}
\usepackage{jheppub} % for details on the use of the package, please
\usepackage[T1]{fontenc} % if needed
\usepackage{color}
\usepackage{ulem}
\usepackage{braket}
\usepackage{mathtools}
\usepackage{graphicx,subfigure,adjustbox}
\usepackage{comment}

\newcommand{\pd}{\partial}

\title
{On catalyzed vacuum decay around a radiating black hole and the crisis of the electroweak vacuum}
\author[a,b]{Takumi Hayashi,}
\author[b]{Kohei Kamada,}
\author[c]{Naritaka Oshita,}
\author[a,b,d,e]{Jun'ichi Yokoyama}

\affiliation[a]{Department of Physics, Graduate School of Science,
The University of Tokyo,\\ Hongo 7-3-1
Bunkyo-ku, Tokyo 113-0033, Japan}
\affiliation[b]{Research Center for the Early Universe (RESCEU), Graduate School of Science,\\ The University of Tokyo, Hongo 7-3-1
Bunkyo-ku, Tokyo 113-0033, Japan}
\affiliation[c]{Perimeter Institute, 31 Caroline St., Waterloo, Ontario, N2L 2Y5, Canada}
\affiliation[d]{Kavli Institute for the Physics and Mathematics of the Universe (Kavli
 IPMU), UTIAS, WPI, The University of Tokyo, Kashiwa, Chiba, 277-8568, Japan}
\affiliation[e]{Trans-scale Quantum Science Institute,\\ The University of Tokyo, Hongo 7-3-1
Bunkyo-ku, Tokyo 113-0033, Japan}
\emailAdd{takumi\_hayashi@resceu.s.u-tokyo.ac.jp}
\emailAdd{kohei.kamada@resceu.s.u-tokyo.ac.jp}
\emailAdd{noshita@perimeterinstitute.ca}
\emailAdd{yokoyama@resceu.s.u-tokyo.ac.jp}

\begin{document}

\begin{flushleft}
RESCEU-10/20 
\end{flushleft}

\abstract{
False vacuum decay is a key feature in quantum field theories and exhibits a distinct signature in the early 
Universe cosmology. It has recently been suggested that 
the false vacuum decay is catalyzed by a black hole (BH), which might cause the catastrophe of the Standard Model Higgs vacuum if primordial BHs are formed in the early Universe.
We investigate vacuum phase transition of a scalar field around a radiating BH with taking into account the effect of Hawking radiation. We find that the vacuum decay rate slightly decreases in the presence of the thermal effect since the scalar potential is stabilized near the horizon. However, the stabilization effect becomes weak at the points sufficiently far from the horizon. Consequently, we find that the decay rate is not significantly changed unless the effective coupling constant of the scalar field to the radiation is extremely large. 
This implies that the change of the potential from the Hawking radiation does not help prevent the Standard Model Higgs
 vacuum decay catalyzed by a BH.
}

\maketitle

\section{Introduction}
Vacuum phase transition is one of the most important phenomenon in the early universe predicted by quantum field theory. If our Universe is in a metastable state, namely a local minimum of a potential, quantum tunneling to the global minimum may take place. It includes a series of physical processes such as vacuum bubble nucleation. 
Such bubbles expand and collide with each other,  which may become an important source of gravitational wave background~\cite{Mazumdar:2018dfl,Caprini:2015zlo}. 
Such metastable vacua often appear in various models of particle physics and string theories~\cite{Peccei:2006as,Branco:2011iw,Kachru:2003aw}, and the most familiar example is the instability of electroweak vacuum~\cite{Arnold:1989cb,Sher:1988mj,Arnold:1991cv,Espinosa:2007qp}. 
The up-to-date measurements of the Higgs mass and top quark mass suggest that 
the effective potential of the Standard Model (SM) Higgs field develops a negative value at the scale higher than $10^{11}$ GeV 
at their median values~\cite{Degrassi:2012ry,Buttazzo:2013uya},  if there are not
any corrections from the physics beyond the SM and quantum gravity. 
If the electroweak vacuum we live were unstable under such a phase transition with a considerable rate, 
our very existence would be in danger. 

The rate of vacuum tunneling associated with such instabilities can be estimated by using the Euclidean path integral technique~\cite{Coleman:1977py,Callan:1977pt}, and subsequently gravitational effects were incorporated in Ref.~\cite{Coleman:1980aw}. Applying it to the case of the Higgs instability, fortunately, the probability that our Universe undergoes the phase transition in the cosmic age is about $O(10^{-600})$~\cite{Branchina:2014rva,Chigusa:2017dux}. 
It is also sufficiently stable against the thermal phase transition regardless of the reheating temperature of the Universe~\cite{Espinosa:2007qp,EliasMiro:2011aa}. 
The inflationary fluctuation is problematic especially for high-scale inflation, 
but several resolution has been proposed~\cite{Kamada:2014ufa,Hook:2014uia,Herranen:2014cua}. 
See also recent discussions on the SM Higgs vacuum stability in the inflationary Universe~\cite{Kohri:2016qqv,Espinosa:2017sgp,Ema:2017loe}.
However, these works assume the homogeneity of the initial vacuum state and the inhomogeneity of the Universe is not taken into account.

It was found that the existence of spatial inhomogeneities greatly changes the result of the tunneling calculation, which was pioneered by Steinhardt \cite{Steinhardt:1981ec} and Hiscock \cite{Hiscock:1987hn}. Hiscock first discussed the bubble nucleation around a non-rotating BH and showed that it can enhance the decay rate~\cite{Hiscock:1987hn}. Gregory {\it et al}. refined and generalized his analysis~\cite{Gregory:2013hja,Burda:2015yfa,Burda:2016mou,Gregory:2020cvy} and pointed out that a very small BH greatly induces the phase transition process (See also~\cite{Canko:2017ebb}). Recently the study of catalyzing effects on cosmological vacuum decay was extended to the cases of spinning BHs \cite{Oshita:2019jan} and other various cosmological impurities \cite{Oshita:2018ptr,Koga:2019mee,Firouzjahi:2020hfq,Oshita:2020ksb}.

Cosmological creation of primordial black holes (PBHs) has been of interest for a long time to 
explain dark matter~\cite{Hawking:1971ei,Carr:1974nx,Carr:1975qj} and recently for the binary BHs~\cite{Sasaki:2016jop} detected by the LIGO and VIRGO~\cite{Abbott:2016blz}. 
The density perturbations in the early universe could also lead to small-mass PBHs which have evaporated by today. Since those PBHs may play the role of catalysts for the Higgs vacuum decay, cosmological parameters relevant to the PBH formation may be constrained by the parameters of the Higgs potential or vice versa \cite{Dai:2019eei}. The footprint of PBH evaporation could remain in stochastic gravitational waves~\cite{Inomata:2020lmk}, which may be detected by near-future GW observations. Therefore, the catalyzing effect of BHs are very important not only in the particle physics but also in the early cosmology.

According to the studies in Ref.~\cite{Gregory:2013hja,Burda:2015yfa,Burda:2016mou}, 
the formation of such small BHs might be dangerous for our Universe since they can induce the SM Higgs vacuum phase transition at the end stage of the evaporation. However, the thermal effects of small BHs could be non-negligible since the Hawking temperature is proportional to the inverse of its mass, and it has been an open question whether the thermal effects of small BHs would stabilize the present Higgs vacuum state or not\footnote{For example, Ref.~\cite{Kohri:2017ybt} argues that the thermal effects of Hawking radiation would stabilize the present Higgs vacuum state and a small BH does not play a role of a catalyst for the Higgs vacuum decay. However, their approach differs from the Euclidean path integral, utilized in~\cite{Gregory:2013hja}, and the stochastic fluctuation of the Higgs field is assumed although the background is Schwarzschild spacetime.}.

In order to investigate this phenomenon in more detail, 
it is important to construct the bounce solution based on 
the effective potential including thermal (quantum) corrections
appropriately. 
Indeed, since we are interested in tiny evaporating BHs, 
we need to use the effective action that describes radiating BHs.
In particular, the Hawking radiation emitted from the BHs might stabilize the Higgs potential or make the Higgs potential barrier high enough to prevent the transition from the false to true vacuum state~\cite{Kohri:2017ybt}. Such a backreaction is quite important when we discuss the phase transition in thermal plasma~\cite{Linde:1981zj,EliasMiro:2011aa}, where quantum fields obtain large vacuum polarization associated with the high temperature. 
In this paper, we take into account this effect on the analysis of the vacuum phase transition around a radiating BH and try to improve the evaluation of the transition rate.

\begin{figure}
\centering
\includegraphics[width=10cm]{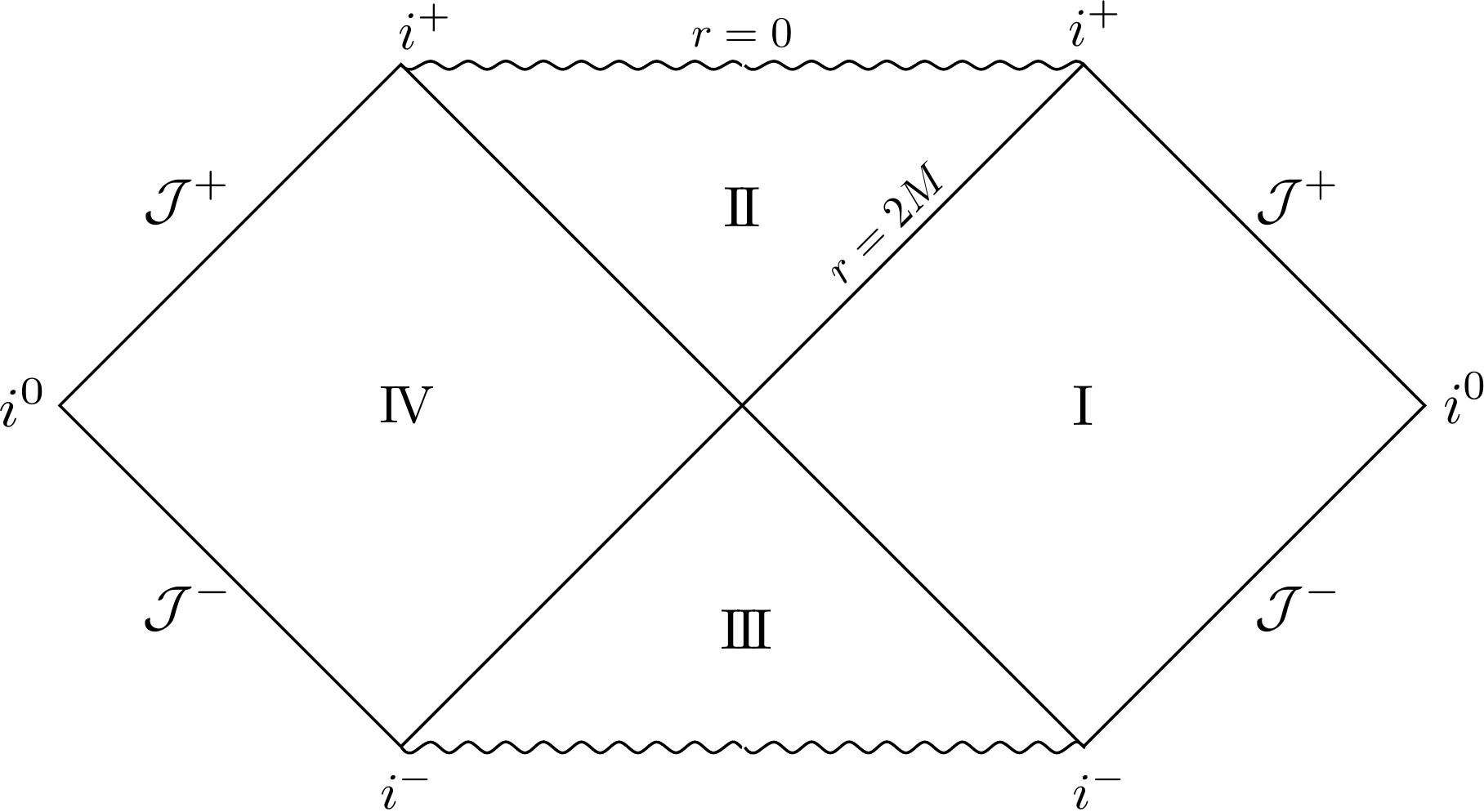}
\caption{The Penrose diagram of maximally extended Schwarzschild spacetime. Region $\rm I$ and $\rm I\hspace{-.1em}I$ are the outside and the inside of the event horizon respectively, and region $\rm I\hspace{-.1em}I\hspace{-.1em}I$ is the white hole interior. Region $\rm I\hspace{-.1em}V$ is the mirror domain of region $\rm I$. Physically relevant regions to a BH formed by gravitational collapse are Region $\rm I$ and $\rm I\hspace{-.1em}I$.}
\label{fig:penrose}
\end{figure}

The crucial point in our analysis is the choice of the vacuum state in the Schwarzschild spacetime. We are interested in the cosmological application of the transition process around a BH. Therefore, we here consider a vacuum state around a gravitationally collapsed BH which has no past horizon that separates the regions $\mathrm{I}$ and $\rm I\hspace{-.1em}I\hspace{-.1em}I$ in the Penrose diagram (Fig.~\ref{fig:penrose}). In this case, the state of quantum fields can be modeled by the Unruh vacuum state~\cite{Unruh:1976db} that leads to outgoing energy flux of vacuum fluctuations and the evaporation process of BHs. On the other hand, the Hartle-Hawking vacuum state~\cite{Hartle:1976tp} describes a thermal equilibrium state around a BH and it does not evaporate\footnote{There is a related work~\cite{Mukaida:2017bgd} that suggests the large enhancement of the transition rate can be interpreted as 
the thermal production of a bubble in the Hawking radiation in the Hartle-Hawking vacuum state (thermal equilibrium state).}. An important feature of the Higgs potential thermally corrected in the Unruh vacuum state is that the thermal correction becomes weaker at a larger distance from the horizon~\cite{Candelas:1980zt}. 
As a result, we will show that the thermal effects do not prevent the catalyst effect of BHs since the Hawking temperature is suppressed near a bubble wall unless there are many light scalar fields $\chi_i$ coupling to the metastable scalar field $\phi$ (e.g. Higgs field) as $ \sim \sum_i \lambda_i \chi_i^2 \phi^2$ with $\sum_i \lambda_i \gg 10^3$, where $\lambda_i$ is a coupling constant.
This general result also holds in the case of the SM Higgs vacuum 
and hence the tiny PBH formation in the early Universe is still a threat of our Universe.

The organization of this paper is as follows. In the next section, we review the vacuum phase transition around a non-radiating BH following Refs.~\cite{Burda:2015yfa,Burda:2016mou} and obtain simple expression of transition rate 
with the thin-wall approximation. In Sec.~\ref{sec3}, we include the effect of Hawking radiation in the potential, calculate the bounce solution, and estimate the transition rate. In Sec.~\ref{sec4}, we apply this result to the case of Higgs vacuum instability and argue that a vacuum bubble nucleation due to the Higgs instability would be catalyzed by a BH even when the thermal effect of Hawking radiation is taken into account. Sec.~\ref{sec5} is devoted to conclusion and discussion. 

We adopt the Planck units for simplicity; the Newton constant $G=1$, Planck mass $m_\mathrm{pl}=1$, and Planck length $\ell_{\rm pl} = 1$. When $G$ explicitly appears in some equations or definitions, it is intended that $G$ denote the original mass dimension of physical quantities.

\section{Static bounce around a ``zero-temperature'' BH}
We start with the review of vacuum phase transition around a Schwarzschild BH following Ref.~\cite{Burda:2016mou}. 
Here we consider a scalar field $\phi$ with a potential that has two minima,  $\phi=\phi_\mathrm{fv}(=0)$ and $\phi=\phi_\mathrm{tv}$, 
where the former is a false vacuum with a vanishing potential energy density and the latter is a true vacuum with a negative potential energy density, 
and evaluate the phase transition rate from the false to true vacuum through the bubble nucleation around the BH. 

\subsection{General discussion on the bounce and vacuum decay rate catalyzed by BH}
Let us consider the following action
\begin {align}
& S = \int_{\mathcal{M}} d^4 x\sqrt {-g}\left (\frac {1} {16\pi} \mathcal{R} - \frac {1} {2} \
(\pd_\mu\phi)^2 - V (\phi) \right) +\frac 1 {8\pi } \int_{\pd\mathcal{M}} K dS
\end {align}
where $\mathcal M$ is the spacetime manifold with a BH, $g$ is the determinant of the (Lorentzian) metric $g_{\mu\nu}$ 
with the convention $(-,+,+,+)$, $\mathcal{R}$ is the Ricci scalar, 
and $K$ is the trace of the extrinsic curvature of the boundary of the spacetime. 
Since we consider the spacetime with a BH horizon, we explicitly include the Gibbons-Hawking-York boundary term. 
Assuming the spherical symmetry and staticity of the final configuration, a general metric is taken to be
\begin{align}
 ds^2  = -f(r)e^{2\delta (r)} dt^2 + f^{-1}(r) dr^2 + 
  r^2 d\Omega_ 2^2, \quad  f  = 1 - \frac {2\mu (r)} {r},
\end {align}
where $d\Omega_2^2$ is the line elements on a unit sphere $S^2$. Solving the Einstein equation, together with the equation of motion for scalar fields, one can obtain the metric functions $\mu(r)$ and $\delta (r)$ and can determine a vacuum bubble configuration. On the other hand, the initial configuration is given by the Schwarzschild metric with $M_+$ being the seed BH mass
\begin{align}
 ds^2 = - \left(1-\frac{2M_+}{r}\right) dt^2 + \left(1-\frac{2M_+}{r}\right)^{-1} dr^2 + 
  r^2 d\Omega_ 2^2. \label{schwarzschildmetric}
\end {align}
%As is shown in Refs.~\cite{Coleman:1977py,Callan:1977pt}, 
In order to estimate the Euclidean action, let us implement the Wick rotation $t \to -i \tau$, and one obtains 
%it is convenient to go to 
the Euclidean spacetime
\begin{align}
 ds_\mathrm{E}^2 = g_{\mathrm{E}\mu\nu} d x^\mu dx^\nu =  fe^{2\delta} d\tau^2 + f^{-1} dr^2 + 
  r^2 d\Omega_ 2^2.
\end {align}
Then one can construct the bounce solutions on the Euclidean background. The existence of a BH reduces the maximal symmetry of bounce to $O(3)$ from $O(4)$. Therefore, we here consider time-independent static $O(3)$ symmetric solutions.
We can still construct time-dependent bounce solutions, but it has been shown that 
the least Euclidean action is given by the static one based on the analysis under the thin-wall condition~\cite{Gregory:2013hja}.
Note that the bounce is static even after analytic continuation to Lorentzian bubble, 
but it is unstable under the perturbations and can easily expand so that the whole system quickly falls down to the (unwanted) 
true vacuum. 
Thus we regard the bubble nucleation rate, evaluated from the bounce action, as the vacuum decay rate of the system. 

The equation of motion for the scalar field and the Einstein equations are
\begin{align}
&f\phi''+f'\phi'+\frac{2}{r}f\phi'+\delta'f\phi'-\frac{\partial V}{\partial\phi}=0,  \label{bounce eq1} \\
&\mu'=4\pi r^2\left(\frac{1}{2}f\phi'^2+V\right), \label{bounce eq2}\\
&\delta'=4\pi r \phi'^2,  \label{bounce eq1'}
\end{align}
where the prime denotes the derivative with respect to $r$. Substituting Eq. (\ref{bounce eq1'}) into Eq. (\ref{bounce eq1}), one obtains
\begin{equation}
f\phi''+f'\phi'+\frac{2}{r}f\phi'+4\pi f r \phi'^3-\frac{\partial V}{\partial\phi}=0. \label{bounce eq1''}
\end{equation}
%Note that $\delta$ has the arbitrariness that is absorbed into the definition of $\tau$ and appear with $\delta'$ in the bounce equations. Thus we do not have to solve the equations for $\delta$ (Eq.~\eqref{bounce eq1'}) but solve the following equation, 
%\begin{equation}
%f\phi''+f'\phi'+\frac{2}{r}f\phi'+4\pi r \phi'^2 f\phi'-\frac{\partial V}{\partial\phi}=0, \label{bounce eq1''}
%\end{equation}
%which is obtained by plugging $\delta'$ in Eq.~\eqref{bounce eq1'} into Eq.~\eqref{bounce eq1}. 
We require that the scalar field is in the false vacuum state at infinity and
%the metric is the one with the 
the asymptotic spacetime is the Schwarzschild spacetime:
\begin{align}
\mu (r)\rightarrow M_+, \quad \phi (r)\rightarrow \phi_\mathrm{fv}, \quad \phi'(r) = 0,  \quad (r \rightarrow\infty),   \label{boundary}
\end{align}
where $M_+$ denotes the initial BH mass before the nucleation. We also impose the following boundary condition at the horizon $r=r_h$
\begin{align}
\mu (r)\rightarrow \mu_-, \quad \phi (r)\rightarrow \phi_0  \quad (r \rightarrow r_h),   \label{boundary'}
\end{align}
where
\begin{align}
\mu_- \equiv \mu(r_h)=\frac{r_h}{2},
\end{align}
and $\mu_-$ and $\phi_0$ are determined by the shooting method so that the obtained solution satisfies the condition \eqref{boundary}.
In order for the solution to avoid the coordinate singularity at the horizon in Eq.~\eqref{bounce eq1}, the following condition should be satisfied at the same time
\begin{align}
\phi'(r_h)=\frac{r_h\frac{\partial V}{\partial\phi}(\phi_0)}{1-8\pi {r_h}^2 V(\phi_0)}.  \label{boundaryphi'}
\end{align}
One can calculate $\delta (r)$ after obtaining $\phi (r)$ from the integration of Eq. (\ref{bounce eq2}) and (\ref{bounce eq1''}), and its integration constant can be absorbed into the scale of $\tau$.
Furthermore, we perform the coordinate transformation to improve the numerical behavior near the horizon as
\begin{align}
r^*=\int \frac{dr}{f(r)}\label{tortoise},
\end{align}
which runs from the horizon $-\infty$ to the spacial infinity $\infty$.
Consequently, the bounce equations Eqs.~\eqref{bounce eq2}, and \eqref{bounce eq1''} become
\begin{align}
&\frac {d^2\phi}{{dr^*}^2}+\frac{2f}r\frac {d\phi}{dr^*}+\frac{4\pi r}{f}\left(\frac {d\phi}{dr^*}\right)^3-f\frac{\pd V}{\pd \phi}=0,\\
&\frac{d\mu}{dr^*}=4\pi r^2\left(\frac1 2\left(\frac {d\phi}{dr^*}\right)^2+fV\right),
\end{align}
and the boundary conditions at the horizon (Eqs.~\eqref{boundary'} and~\eqref{boundaryphi'}) become

\begin{align}
\mu (r^*)\rightarrow \mu_-, \quad \phi (r^*)\rightarrow \phi_0,\quad d\phi/ dr^{\ast} \to 0\quad (r^* \rightarrow -\infty).
\end{align}

The exponent of vacuum decay rate can be evaluated by the difference between the Euclidean action $S_E$ of the bounce and of the false vacuum solution before the transition, and the full form of the decay rate can be obtained as \cite{Gregory:2013hja}
\begin{align}
\Gamma_\mathrm{D}  \sim \sqrt{ \frac B {2\pi}} M_+^{-1} \exp\left(-B\right),\quad \text{with} \quad  B=S_E[g_E,\phi]-S_E[g_\mathrm{ESch},\phi_\mathrm{fv}] \label{transition rate}
\end{align}
where $M_+$ is initial BH mass and $g_\mathrm{Esch}$ denotes the Euclideanized Schwarzschild metric, see Eq.~\eqref{schwarzschildmetric}. 
$(\phi,g_E)$ represents the bounce solution obtained by solving the bounce equations. 
The dimensionless factor $\sqrt{B/2\pi}$ comes from the normalization factor of the zero mode around the bounce associated with the time translation symmetry. The dimensionful prefactor $M_+^{-1}$ is taken from the typical energy scale of the transition process~\cite{Gregory:2013hja}. For the static bounce solution, the bulk part of the action vanishes due to the Hamiltonian constraint and only the boundary part, which comes from the Gibbons-Hawking-York term, contributes to the action~\cite{Gregory:2013hja}. 
The boundary contribution reduces to $(-1)$ times the Bekenstein-Hawking entropy that can be derived without tuning the period of the Euclidean time (even with the conical singularity)
\begin{align}
S_E[g_E,\phi]=- \frac{\mathcal{A}}{4},
\end{align}
where $\mathcal{A}$ is the horizon area~\cite{Gregory:2013hja}. Consequently the bounce action is determined to be 
\begin{align}
B=4 \pi (M_+^2-\mu_-^2). \label{bounce action}
\end{align}
Since this expression generally holds for static bounce solutions, 
we will use the expression in the next section where we consider static bounce solutions with radiating BHs.

\subsection{Thin-wall approximation}

In principle we need to numerically solve the bounce equations (\ref{bounce eq1}), (\ref{bounce eq2}) and (\ref{bounce eq1'}) for a given potential
to see how the vacuum decay rate changes in the presence of a BH. 
However, when the thin-wall approximation is applicable, we can give an analytic investigation
to show some qualitative and quantitative features of the BH catalyst effect. 

In the thin-wall approximation, one can suppose that an infinitely-thin wall separates two different spacetimes and the scalar field can be represented by the step function at the wall. The system is, then, characterized by the tension of the wall $\sigma$, the bubble radius $R$, and the interior vacuum energy density, characterized by anti-de Sitter (AdS) radius $l \equiv \sqrt{3/|8\pi V (\phi_\mathrm{tv})|}$, where the false vacuum energy density is assumed to be zero. The details of the effective potential barrier is ``coarse-grained'' in the tension parameter $\sigma$ as
\begin{align} 
\sigma=\int^{\phi_\mathrm{tv}}_{\phi_\mathrm{fv}}d\phi\sqrt{2 V(\phi)} \sim V_\mathrm{barrier}^{1/2} \Delta \phi,\quad \Delta \phi \equiv |\phi_\mathrm{tv}-\phi_\mathrm{fv}|, 
\end{align} 
where $V_\mathrm{barrier}$ is the maximum value of effective potential between the false and true vacuum, 
and we suppose it is much larger than
the potential energy difference,  $V_\mathrm{barrier} \gg \Delta V \equiv |V(\phi_{\rm fv}) - V(\phi_{\rm tv})|$. 
Thin-wall approximation is appropriate if the 
bubble-wall thickness is thinner than any relevant length scales, e.g. the bubble radius. While the bubble radius is determined by the balance between the energy contributions from the bulk $E_\mathrm{bulk}$ and the wall $E_\mathrm{wall}$. For a spherical bubble, we roughly estimate
\begin{equation}
|E_\mathrm{bulk}| \sim \frac{4\pi}{3} R^3 \Delta V, \quad E_\mathrm{wall} \sim 4 \pi R^{2} \sigma \simeq 4 \pi V_\mathrm{barrier}^{1/2} \Delta \phi R^2, 
\end{equation}
which give the bubble radius in terms of the potential parameters,
\begin{equation}
    R \sim \frac{V_\mathrm{barrier}^{1/2} \Delta \phi}{\Delta V}. \label{estimateR}
\end{equation}
On the other hand the wall thickness is determined by the curvature of the potential top at $\phi = \phi_{\rm barrier}$, $l_\mathrm{wall} \sim m_\mathrm{barrier}^{-1} \equiv \sqrt{V''(\phi_{\rm barrier})}$.
Thus we see that the thin-wall approximation is appropriate when the energy difference between the 
false and true vacuum, $\Delta V$, is sufficiently small, 
\begin{equation}
    \Delta V \ll  V_\mathrm{barrier}^{1/2} m_\mathrm{barrier} \Delta \phi. \label{thinwallcond}
\end{equation}

In the following, we assume that Eq.~(\ref{thinwallcond}) holds and calculate the dynamics of a thin-wall bubble which is consistent with the Einstein equation. To this end, we solve the Israel junction condition~\cite{Israel:1966rt} with the $O(3)$ symmetry:
\begin{equation}
K_{ab}^{(+)} - K_{ab}^{(-)} = 8 \pi G \left(S_{ab}- \frac{1}{2} h_{ab} S\right),
\end{equation}
where $K_{ab}^{(+/-)}$ is the extrinsic curvature outside/inside the bubble wall, $h_{ab}$ is the induced metric on the wall, and $S_{ab} = -\sigma h_{ab}$ is the energy momentum tensor of the wall. The trajectory of the bubble is given by
\begin{align} 
X^{\mu}_\pm=(\tau_\pm(\eta),R(\eta),\theta,\phi),
\end{align}
where $\eta$ is Euclidean proper time and $\tau_{+/-}$ is the Schwarzschild time for the exterior/interior spacetime. Outside the bubble we take the scalar field is at the false vacuum and the metric is the Schwarzschild metric, 
\begin{equation}
\phi = \phi_\mathrm{fv}, \quad ds_{\mathrm{E}+}^2=f_+ d\tau_+^2+f_+^{-1}dr_+^2+r^2d\Omega_2^2,\quad \text{with} \quad f_+ (r)=1-\frac{2M_+}{r}, 
\end{equation}
whereas inside the bubble we take the scalar field is at the true vacuum and the metric is the Schwarzschild-AdS metric,
\begin{equation}
\phi=\phi_\mathrm{tv}, \quad ds_{\mathrm{E}-}^2=f_- d\tau_-^2+f_-^{-1}dr_-^2+r^2d\Omega_2^2,\quad \text{with} \quad f_- (r)=1-\frac{2M_-}{r}+\frac{r^2}{l^2},  
\end{equation}
where $M_{+}$ and $M_-$ are the BH masses before and after the bubble nucleation, respectively, and we set $r_+ = r_- = R$ at the bubble wall.  
Note that the horizon and the BH mass inside the bubble are related as 
\begin{align}
M_-=\frac{4}{l^2}\mu_-^3+\mu_-, \quad \mu_- = r_h/2. \label{horizon mass}
\end{align}

Then the $(\theta, \theta)$-component of the Israel junction condition reduces to
\begin{align} 
f_+(R)\dot{\tau}_+-f_-(R)\dot{\tau}_-=-\frac{\overline{\sigma}R}{2},  \quad (\overline{\sigma}\equiv 8\pi\sigma),  \label{junction}
\end{align} 
where the dot represents the derivative with respect to $\eta$. 
Imposing the condition that the magnitude of the wall four-velocity to be the unity, one obtains
\begin{align} 
f_\pm(R) \dot{\tau}^2_\pm+\frac{\dot{R}^2}{f_\pm(R)}=1. \label{proper}
\end{align} 
Using Eq. (\ref{proper}), the Israel junction condition (\ref{junction}) reduces to
\begin{align} 
\dot{R}^2=1-\left(\frac{\overline{\sigma}^2}{16}-\frac{1}{2 l^2}+\frac{1}{\overline{\sigma}^2 l^4 }\right)R^2-\left(M_++M_-+\frac{4(M_+-M_-)}{\overline{\sigma}^2l^2}\right)\frac{1}{R}-\frac{4(M_+-M_-)^2}{\overline{\sigma}^2 R^4}. 
\end{align} 
By redefining the  variables as
\begin{equation}
{\tilde R} = \alpha R, \quad {\tilde \tau} = \alpha \tau, \quad {\tilde \eta} = \alpha \eta, \quad \text{with} \quad \alpha \equiv \frac{1-\overline{\sigma}^2 l^2/4}{\overline{\sigma}l^2}, 
\end{equation}
the equation for the wall position is rewritten as
\begin{align} 
\frac{1}{2}\left(\frac{d\tilde{R}}{d\tilde{\lambda}}\right)^2+U(\tilde{R})=0,\quad 
2U(\tilde{R})\equiv-1+\left(\tilde{R}+\frac{k_2}{{\tilde{R}}^2}\right)^2+\frac{k_1}{\tilde{R}}.
\end{align} 
where
\begin{align}
 &k_1=2\alpha M_+ , \quad 
  k_2=\frac{2\alpha^2(M_+-M_-)}{\overline{\sigma}}.\label{k1(M)}
\end{align}
Here we consider the case $\overline{\sigma} l <1/2$. 
Indeed, it was shown that there is no any solutions that satisfy an appropriate junction condition 
for $\overline{\sigma} l >1/2$~\cite{Burda:2015yfa}. 

The $O(3)$ static solution, which gives the highest decay rate, are obtained by requiring that there are parameter sets satisfying $U(\tilde R)=U'(\tilde R)=0$.   
This determines the relation between the parameters $k_1$ and $k_2$ 
\begin{align} 
&k_1=k_1^*(k_2)\equiv
    -2k_2\mp\frac{2}{9}\sqrt{1+81 k_2^2+Q_+-Q_-}\quad  \text{for}\quad  k_2\lessgtr-\frac{2}{3\sqrt{3}} \label{k1(k2)}\\
&\text{with} \ Q_\pm=\left(\pm\left(1+5(27k_2)^2-\frac{(27k_2)^4}{2}\right)+\frac{27k_2}{2}\left((27k_2)^2+4\right)^{\frac3 2} \right)^{\frac{1}{3}},  \label{k1k2rel}
\end{align}
as well as the position of the wall, 
\begin{align}
&{\tilde R}={\tilde R}_*\equiv 2^{-\frac2 3}\left(k_1^*+2k_2+\sqrt{k_1^*+4k_1^*k_2+36k_2^2}\right)^{\frac1 3}.
\end{align}
Note that $k_1^*$ is positive for $k_2<4/27$. 

From Eq.~\eqref{k1(k2)} we can determine $M_-$ (and hence $\mu_-$ from Eq.~\eqref{horizon mass}) 
in terms of $\sigma$, $l$, and $M_+$. 
The bounce action is then determined by Eq.~\eqref{bounce action}. 
Note that the effective potential governing the position of the wall is concave, $W''(\tilde R)<0$, in the Lorentzian picture, $W(\tilde R) \equiv -U(\tilde R)$. Therefore, the static bubble is unstable and it may eventually collapse or expand after the nucleation. If the Hawking radiation that has outgoing flux interacts with the bubble wall, it may push the wall outward and the bubble would expand.
Thus, as we have mentioned in the above, we expect that the nucleated bubbles are 
likely to expand to the spatial infinity, which is the catastrophe of the Universe, 
and hence we treat the bubble nucleation rate as the vacuum decay rate.

It is instructive to see the large and small-$M_+$ limits of the solution. 
For $k_2 \ll -1$ (large-$M_+$ limit), we see $k_1^*\gg 1$ or $M_+ \gg 1/2\alpha$ so that
\begin{equation}
k_1^*  \simeq  (-k_2)^{1/3}+\frac1{36(-k_2)^{1/3}} +{\cal O} \left( (-k_2)^{-2/3} \right), \quad R_* \simeq 2 M_+ \left(1+ \frac{1}{144 \alpha^2 M_+^2}\right),    \label{apk1(k2)}
\end{equation}
which implies that  the bubble is created near the horizon. 
From Eqs.~\eqref{bounce action}, ~\eqref{horizon mass},~\eqref{k1(M)}, and \eqref{apk1(k2)}, then, we obtain 
the horizon mass inside the bubble and the bounce action as 
\begin{align}
&\mu_-\simeq (1-\overline\sigma^2l^2/4)^{\frac1 3}M_+,\\
&B\ \simeq 4\pi\left(1-(1-\overline\sigma^2l^2/4)^{\frac2 3}\right)M_+^2\label{B1}, 
\end{align}
at the first order in $M_+$.

\begin{figure}
    \centering
    \includegraphics[width=12cm]{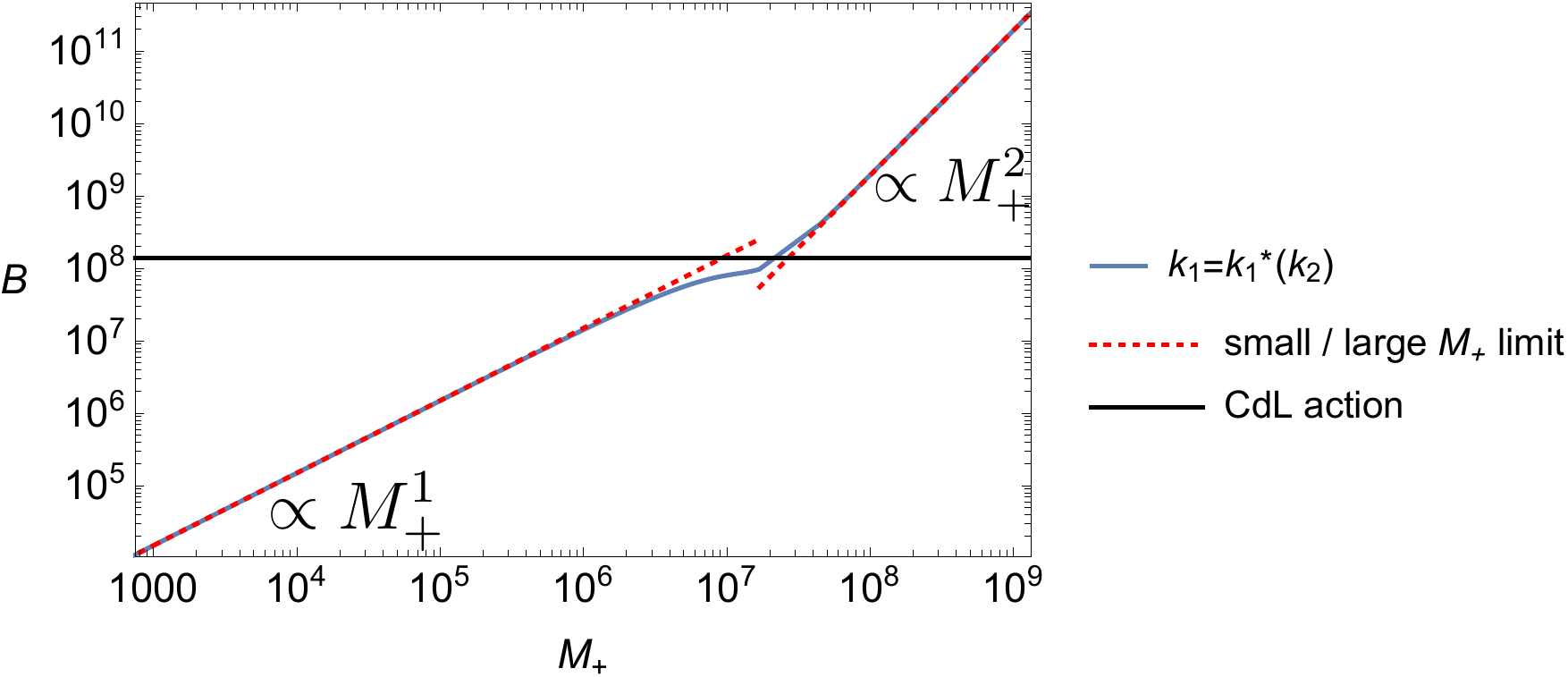}
    \caption{The bounce action dependence on initial BH mass $M_+$ for $\overline\sigma = 3.42 \times 10^{-15}$ 
and $l = 9.36 \times 10^{10}$ which corresponds to the  potential parameters (Eq.~\eqref{potential}) $m=4.08\times10^{-6}$, $g=2.13$, $\lambda=1$.  The blue line represents the bounce action obtained from the condition $k_1=k_1^*(k_2)$ (Eq.~\eqref{k1(k2)}), 
and the red dotted lines the asymptotic forms given by Eqs.~\eqref{B1} and \eqref{B2}. Black solid line means the CDL bounce action given by Eq.~\eqref{BCDL}. The bounce action of the $O(3)\times U(1)$ bubble  become significantly smaller than the CDL's one as $M_+\ll10^7$.
    \label{fig:k1(k2)}}
\end{figure}

For $k_2 \sim 4/27$ (small-$M_+$ limit), we see $k_1\ll 1$ or $M_+ \ll 1/2\alpha$, which leads to 
\begin{equation}
k_1^* \simeq 3\left(\frac{4}{27}-k_2\right), \quad  R\simeq \frac{2}{3 \alpha} - \frac{M_+}{2}, \label{k1(k2)2}
%2M_+ \left(\frac{1}{3\alpha M_+}-\frac{1}{4} \right),
\end{equation}
which means that the bubble wall is placed far from the horizon, and the configuration is close to 
an static $O(3)$ bubble without a BH. Combining Eq.~\eqref{k1(k2)2} with Eqs.~\eqref{bounce action}, ~\eqref{horizon mass}, and~\eqref{k1(M)}, we obtain the asymptotic expressions
\begin{align} 
&\mu_-\simeq M_+-\frac{2\overline\sigma^3l^4}{27(1-\overline{\sigma}^2 l^2/4)^2}, \\
&B\ \simeq \frac{16\pi \overline\sigma^3l^4}{27(1-\overline{\sigma}^2 l^2/4)^2}M_++4\pi \left(\frac{2\overline\sigma^3l^4}{27(1-\overline{\sigma}^2 l^2/4)^2}\right)^2. \label{B2}
\end{align}
For $\overline \sigma l \ll 1$, when the gravitational effect is  small, the bounce action is 
further approximated as 
\begin{equation}
B \simeq \frac{16\pi}{27}\overline\sigma^3 l^4 M_+. \label{smallmbounceaction} 
\end{equation}
Figure~\ref{fig:k1(k2)} shows the bounce action as a function of the BH mass before the bubble nucleation $M_+$
with $\sigma = 3.42 \times 10^{-15}$ 
and $l = 9.36 \times 10^{10}$.

Since we are interested in the bubble nucleation catalyzed by radiating BHs,
the bubble nucleation rate evaluated in the above is meaningful only if it is larger than the BH evaporation rate~\cite{MacGibbon:1990zk,MacGibbon:1991tj},  

\begin{align}
\Gamma_\mathrm{H} \equiv \dot{M}/M  \simeq 7.5 \times 10^{-5} g_H  M_+^{-3}, \label{evaporation rate}
\end{align}
where $g_H$ is the  effective degrees of freedom emitted as the Hawking radiation, 
\begin{align}
g_H\simeq\sum_i g_{s_i}, \quad \mathrm{with}\quad g_s=\begin{cases}
1 & s=0,\\
0.55 &s=\frac12,\\
0.22 &s=1,\\
0.003 &s=2. 
\end{cases}
\label{g*}
\end{align}
where $i$ denotes the particle species whose mass is lighter than Hawking temperature, and $s$ is the spin of the particle. Note that $g_H\simeq60$ is obtained in the case of SM for a small BH satisfying $M\ll 10^{17}$.

Since the bounce action increases in proportion to the BH mass before the phase transition, $M_+$, 
the bubble nucleation rate becomes exponentially smaller for larger $M_+$, see Eq.~\eqref{transition rate}. 
Comparing it with Eq.~\eqref{evaporation rate}, which shows the decay of the BH evaporation rate 
proportional to a power law of $M_+$, we see that for smaller $M_+$, the bubble nucleation rate is larger 
than the BH evaporation rate. 
For example, for $\sigma\sim 10^{-15}, l\sim 10^{10}$, and $g_H\sim 60$, 
$\Gamma_\mathrm{H}<\Gamma_\mathrm{D}$ holds for $M_+ \lesssim 10^7$. 
Though this discussion is limited to the thin-wall case, it has been found that 
even in the case when the thin-wall approximation does not hold (see Eq.~\eqref{thinwallcond}), 
$\Gamma_\mathrm{H}<\Gamma_\mathrm{D}$ is likely to be satisfied for small $M_+$~\cite{Burda:2016mou}.   
Note that for $M_+ <1$, the BH radius becomes smaller than the Planck scale so that the (semi)classical analysis of gravity breaks down and hence we do not consider such a case. We here assume that if $\Gamma_\mathrm{H}<\Gamma_\mathrm{D}$ holds at 
$M_+=1$, the Planck-mass BH catalyzes the phase transition to the AdS vacuum.

We can see how the presence of BH changes the vacuum decay rate by comparing it with the CdL tunneling rate. 
In the thin-wall approximation, the bounce action is evaluated as~\cite{Coleman:1980aw}

\begin{align}
B_\mathrm{CDL}=\frac{\pi \overline\sigma^4l^6}{16G(1-\overline\sigma^2 l^2/4)}.\label{BCDL}
\end{align}
The bounce action around a small BH Eq.~\eqref{B2}, which is of our interest as seen in the above, 
is less than the CDL bounce action for $1 < M_+ \ll \alpha^{-1}$.

\subsection{Connecting the parameters of thin-wall bubbles with the effective potential}
The analysis of thin-wall bubbles with Israel junction condition uses some parameters characterizing a vacuum bubble, e.g. $\sigma$ and $l$ in the previous subsection. These parameters include some details of an effective potential of scalar field. We here clarify the relation between the thin-wall parameters and the effective potential.
Let us consider the following toy potential of scalar field,
\begin{align}
V(\phi)= \frac{m^2}{2} \phi^2 -\sqrt{\frac{\lambda}{6}} \frac{g m}{3} \phi^3+\frac{\lambda}{4!} \phi^4 \label{potential},
\end{align}
It has a true vacuum at $\phi_\mathrm{tv}\simeq 2 \sqrt{3} m/\sqrt{\lambda}$ for $g> g_0=3/\sqrt 2$ 
with $V(\phi_\mathrm{tv}) = -4 \sqrt{2} (m^4/\lambda) \Delta g$
to the first order in $\Delta g \equiv g-g_0$, 
and the false vacuum at the origin $\phi_\mathrm{fv} = 0$ with $V(\phi_\mathrm{fv}) = 0$. 
The potential maximum between true and false vacua is given by $V_\mathrm{barrier} = (3/8) m^4/\lambda$. 
Thus $V_\mathrm{barrier} \gg \Delta V$ is satisfied for $\Delta g\ll \sqrt{3+\sqrt{3}} - 3/\sqrt{2} \simeq  0.05$, 
in which case the thin-wall approximation is appropriate, see Eq.~\eqref{thinwallcond}. 
In the following analysis, we use the potential Eq.~\eqref{potential}
with $(m, g, \lambda)$ being independent parameters that fixes the potential.

\begin{figure}
\centering
\includegraphics[width=12cm]{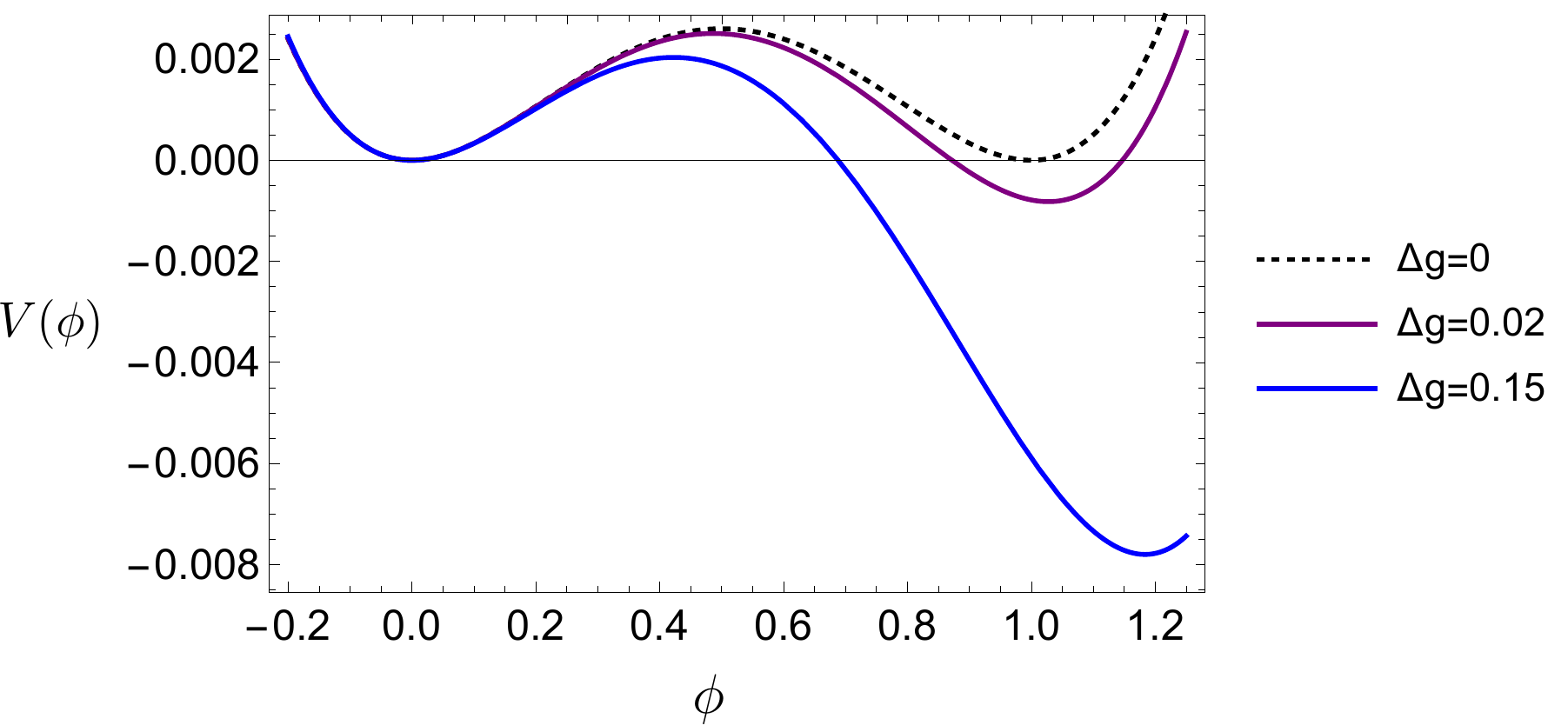}
\caption{Potential shape for $m=1/2\sqrt{3}$, $\lambda=1$. The dashed line represents $\Delta g=0$ for which the two vacua degenerate, and the potentials with the purple and blue lines lead to a thin-wall and thick-wall bubbles, respectively.}
\label{fig:potential}
\end{figure}

Next let us evaluate the thin-wall parameters and results in terms of the potential parameters with $V(\phi)$ Eq.~\eqref{potential}. 
The tension and the AdS radius are given by
\begin{align} 
\overline\sigma=\frac{16\pi m^3}{\lambda},\quad  l=\sqrt{\frac{3}{32 \sqrt 2 \pi}}\sqrt{\frac{\lambda}{\Delta g}}\left(\frac{1}{m}\right)^{2}. 
\end{align} 
Then we obtain $\overline\sigma l\sim (m/\sqrt{\lambda})/\sqrt{\Delta g}\sim \phi_\mathrm{tv}/\sqrt{\Delta g}$, and the condition $\overline \sigma l \ll 1$ is 
satisfied for sub-Planckian false vacuum $\phi_\mathrm{tv} \ll 1$ with $\Delta g \sim {\cal O} (0.1)$. In Ref.~\cite{Burda:2015yfa} it is argued that for $\overline\sigma l > 1/2$ there is no static $O(3)$ bubble solution, which should satisfy $f_\pm\dot\tau_\pm>0$. Now the physical meaning of this condition is clear. 
For  $\overline\sigma l \geq {\cal O} (1)$, the two vacua are degenerated 
too much $\Delta g \simeq \phi_\mathrm{tv}^2 \ll 1$, 
gravitational back reaction prevents the system from 
having a bounce solution.

The bounce action and solutions for $\overline \sigma l \ll 1$ are written as
\begin{align} 
& \alpha \simeq \frac{2\sqrt{2}}{3} m \Delta g,  \\
&\mu_-\simeq  
\begin{cases}
   \left(1- \dfrac{\sqrt{2}\pi m^2}{\lambda \Delta g}\right)M_+&  (M_+\gg1/(2\alpha))\\
   M_+- \dfrac{4 \pi m}{3 \lambda \Delta g^2 } &   (M_+\ll1/(2\alpha))
  \end{cases},\\
 &B\  \simeq  
\begin{cases}
    \dfrac{8\sqrt{2}\pi^2 m^2}{\lambda \Delta g}   M_+^2&  (M_+\gg1/(2\alpha))\\
    \dfrac{32\pi^2 m }{3\lambda \Delta g^2} M_+ &   (M_+\ll1/(2\alpha))
  \end{cases}\label{BofM0},\\
& R\ \simeq \begin{cases} 
2 M_+ \left(1+\dfrac{1}{128 m^2 \Delta g^2 M_+^2}\right)&  (M_+\gg1/(2\alpha))\\
2M_+ \left( \dfrac{1}{2 \sqrt{2} m \Delta g M_+}- \dfrac{1}{4}\right)&   (M_+\ll1/(2\alpha))
\end{cases}, \label{radius}
\end{align} 
which will be useful for the investigation in the next section. The bubble radius (Eq.~\eqref{radius}) is further rewritten as
\begin{equation}
R\simeq \frac{1}{\sqrt{2} m \Delta g } \quad  \text{for} \quad M_+\ll (m \Delta g)^{-1},  \label{radiuslarge}
\end{equation}
for the future use.

\section{Static bounce around a radiating BH} \label{sec3}

While the evaluation in the previous section based on Refs.~\cite{Burda:2015yfa,Burda:2016mou} 
explicitly shows the amplification of the vacuum decay rate due to the seed BH for small mass, 
the bounce solution is obtained from the tree-level action. 
Since a BH emits the high-temperature Hawking radiation, one might expect 
that it stabilizes the scalar potential so that the vacuum decay rate is instead reduced~\cite{Kohri:2017ybt}. 
In order to clarify which effect is dominant, it is straightforward to construct the bounce solution 
and evaluate the decay rate with the effective potential that takes into account the thermal effect from 
the Hawking radiation.
This procedure is similar to the case of the evaluation of usual thermal phase transition
and can be understood as the next-to-minimal order correction 
to the one in Ref.~\cite{Burda:2015yfa,Burda:2016mou}.

We here take into account the vacuum polarization effect of the Hawking radiation and calculate the bounce action with a thermal effective potential. Since we are interested in the spacetime of gravitationally collapsed BHs, which consists of the Region I and $\rm I\hspace{-.1em}I$ in the Penrose diagram (Fig. \ref{fig:penrose}), the state of quantum fields can be modeled by the Unruh vacuum state that gives the outgoing thermal radiation from the vicinity of the future horizon. Since it leads to the inhomogeneous temperature distribution around a BH, it is difficult to carry out the full 1-loop calculation with path integral formalism\footnote{In the case of the Hartle-Hawking vacuum, the full 1-loop effective potential is calculated by performing the path integral in an analogous way of the thermal field theory, with imposing the thermal periodic boundary condition on the Euclidean time. See Ref.~\cite{Flachi:2010yz,Flachi:2011sx} for details.}. Thus, we will only include the thermal mass, which would be the leading contribution, and neglect $O(\lambda^2)$ terms to yield the following expression~\cite{Moss:1984zf}:
\begin{align} 
\Gamma_E[\phi]=\int \sqrt{g_E}d^4x_E \left(\frac 1 2 (\pd_{\mu E}\phi)^2 +V(\phi)+\frac \lambda 4 \braket{U|\phi^2(x_E)|U}\phi^2 \right)+O(\lambda^2), \label{effective action}
\end{align} 
where $\braket{U|\phi^2(x_E)|U}$ is the vacuum polarization with the Unruh vacuum state $|U\rangle$. 
In order to evaluate the vacuum polarization, we adopt the renormalized one for the massless scalar field~\cite{Candelas:1980zt}, which has the asymptotic form as~\cite{Candelas:1980zt,Moss:1984zf}
\begin{align}
\braket{U|\phi^2(x_E)|U}&=\begin{dcases}
\frac 1 {192\pi^2M^2} -\frac1{8\pi^2r^2}\int_0^\infty d\omega\frac{\sum_\ell(2\ell+1)|B_\ell(\omega)|^2 }{\omega(e^{\omega/T_H}-1)} &(r\sim 2M)\\
\frac1{8\pi^2r^2}\int_0^\infty d\omega\frac{\sum_\ell(2\ell+1)|B_\ell(\omega)|^2 }{\omega(e^{\omega/T_H}-1)}&(r\gg 2M)
\end{dcases}\notag \\
& \simeq \begin{dcases}\frac1 {256\pi^2M^2}   &(r\sim 2M) \\
\frac1{192\pi^2r^2}  &  (r\gg 2M) \label{vacuum polarization}, 
\end{dcases}
\end{align}
where $\omega$ is a frequency of Hawking particle, $\ell$ is the angular mode, $M$ is the mass of the Schwarzschild BH, $T_H=1/8\pi M$ is the Hawking temperature, and $B_\ell(\omega)$ is defined as~\cite{Page:1976df} 
\begin{align}
B_\ell(\omega)\sim\left[\frac{\ell!^2}{(2\ell)!(2\ell+1)!!}\right]^2\prod^\ell_{m=1}\left[1+\left(\frac{\omega}{m\kappa}\right)\right]\frac{2\omega}{\kappa}\left(\frac{\omega}{2\kappa}\right)^{2\ell+1}, \quad \text{with} \quad \kappa \equiv 2\pi T_H. 
\end{align}
Here we neglect the masses of Hawking particles, since at the leading order the massless approximation is a good 
approximation for the high temperature regime, which is the case of our interest. For the computational convenience we adopt the values at $r\gg2M$. This treatment is not problematic since the thermal correction of the transition rate becomes important when the BH temperature is higher than the inverse of bubble radius, i.e. when the bubble radius is larger than the BH radius. The size of bubble is governed by the typical scale of phase transition, as we will see later. Moreover, the values of the two-point function near the horizon, $1/256\pi^2 M^2$, is not significantly different from the values extrapolated from the asymptotic expression in Eq.~\eqref{vacuum polarization}. In summary,  the effective potential is written with the radius-dependent thermal mass as
\begin{align} 
V_\mathrm{eff}(\phi,r)\sim V(\phi)+\frac{\lambda}{768\pi^2 r^2}\phi^2. \label{thermalpot}
\end{align} 

We here neglect the change of the spacetime caused by the vacuum bubble nucleation. This is valid when the BH mass inside the bubble, $\mu_{\mathrm{th-}}$, is close to the BH mass outside the bubble, $M_+$, and $GM_+/l \ll 1$ holds. We shall consider the case where this condition is satisfied. Note that at $r\gg2M$ the vacuum polarization Eq.~\eqref{vacuum polarization} is independent of the BH mass $M$, and hence the asymptotic behavior of the effective potential at $r\gg 2M$ seems to be insensitive to the change of the BH mass. If that is so, then our calculation may be extendable to the situation where the seed BH disappears due to a phase transition \cite{Gregory:2013hja}.

So far we have considered the case where the system contains only one scalar field that experiences the phase transition. 
However, we can consider the case where there are many other fields, denoted as $\chi_i$, which couple to the scalar field, say,  
\begin{align}\displaystyle
V_\mathrm{int}=\sum_i \frac{\lambda_i}{4} \chi_i^2\phi^2. 
\end{align}
In such a case, the 1-loop effective potential is approximated as
\begin{align} 
V_\mathrm{eff}(\phi,r)\simeq V(\phi)+\frac{\tilde\lambda}{4}\braket{\hat{\chi}^2}\phi^2 \simeq V(\phi)+\frac{{\tilde \lambda}}{768\pi^2 r^2}\phi^2, \ \tilde\lambda=\sum_i \lambda_i,  \label{effectivepotential}
\end{align}
where the effective coupling constant ${\tilde \lambda}$ is a parameter independent of $\lambda$ in $V(\phi)$. 
In particular, if the number of species coupled to $\phi$ is large, we extrapolate the above expression to ${\tilde \lambda} \gg 4 \pi$, beyond the validity of the perturbative analysis.

\subsection{Bounce solution around a radiating BH}
Since the effective potential of the scalar field is lifted by the thermal effect, the false vacuum can be stabilized near the horizon. Therefore, it is expected that the decay rate is reduced or the bounce solution does not exist due to the stabilization. We here investigate the thermal effects on the bounce solutions and our methodology is presented below.

\begin{figure}
\centering
\includegraphics[width=10cm]{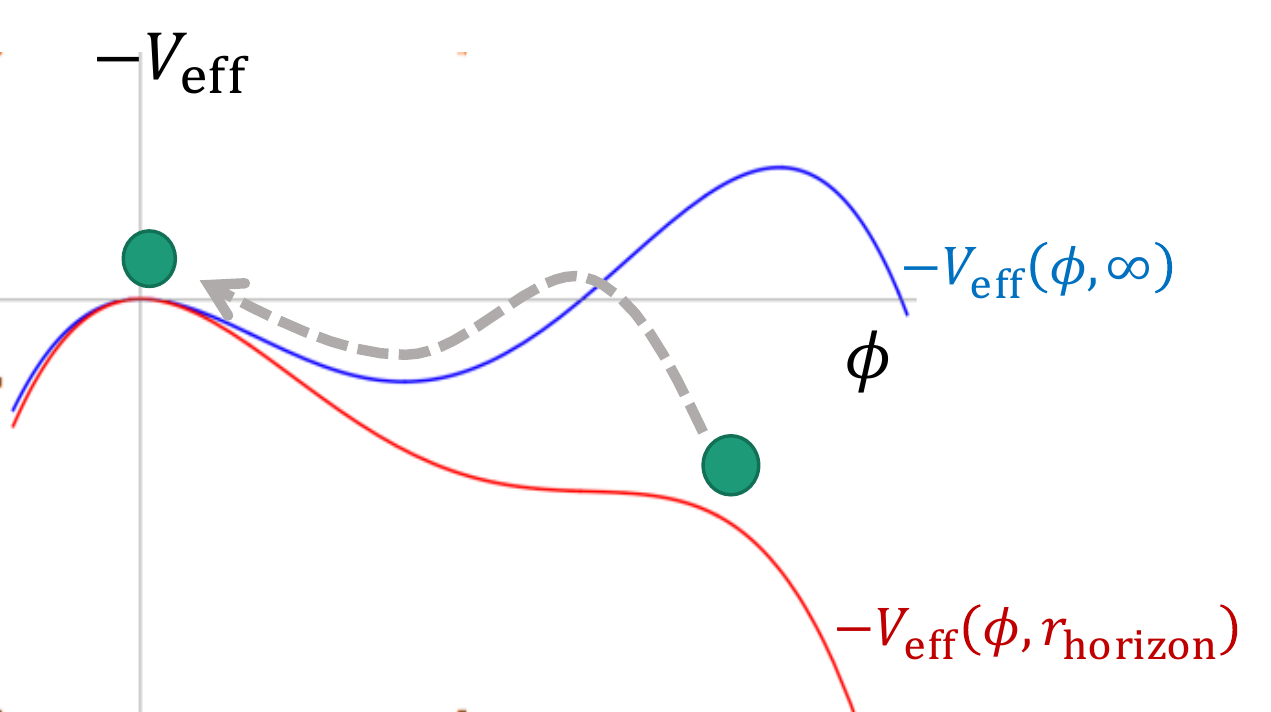}
\caption{The Euclidean dynamics of the scalar field is shown. 
The red line represents the flipped effective potential at the horizon, which is stabilized by the thermal mass and the true vacuum is absent. At a distant region, the flipped effective potential is lifted as the thermal effect becomes weaker (blue line). Therefore, even if the true vacuum is absent near the horizon, a bounce solution can be constructed by using the shooting method (gray dashed line).
}\label{flipped potential}
\end{figure}

Here we consider the static $O(3)$ bounce solution that may give the least Euclidean action and calculate the equation of motion of the scalar field and the Einstein equations:
\begin{align}
&f\phi''+f'\phi'+\frac{2}{r}f\phi'+4\pi r f\phi'^3-\frac{\partial V(\phi)}{\partial\phi}-\frac{\tilde \lambda}{384 \pi^2 r^2}\phi =0, \label{bounce eq3} \\
&\mu'=4\pi r^2\left(\frac{1}{2}f\phi'^2+V(\phi)+\frac{\tilde \lambda}{768 \pi^2 r^2} \phi^2\right).  \label{bounce eq4}
\end{align}
They are obtained by replacing the effective potential $V$ in Eqs. (\ref{bounce eq1}) and (\ref{bounce eq2}) with $V_{\rm eff}$ Eq. (\ref{effectivepotential}). 
We then take the boundary conditions at the spatial infinity as the same as Eq.~\eqref{boundary}. In general, the thermal correction to the potential leads to a corrected remnant mass $\mu_{\mathrm{th}-}$ and horizon radius $r_{h,\mathrm{th}}$, and the boundary condition at the horizon is expressed as 
\begin{equation}
\mu(r_{h,\mathrm{th}})=\mu_{\mathrm{th}-},\quad r_{h,\mathrm{th}}=2\mu_{\mathrm{th}-}. \\    
\end{equation}
The change of the bounce solution can be understood qualitatively as follows. 
As is well known, the bounce equation can be understood as the scalar field dynamics 
with $r$ being the ``time'' coordinate in the flipped potential $-V(\phi)$ 
from the horizon to infinity. 
Without thermal corrections, the flipped potential has two local maxima with $- V(\phi_\mathrm{tv}) > - V(\phi_\mathrm{fv})$. 
The field starts to fall off from the point near the true vacuum to the false vacuum just once $r$ deviates from the horizon. On the other hand, with thermal corrections around a sufficiently small BH, the potential is lifted around the horizon 
so that $-V_\mathrm{eff}(\phi_\mathrm{tv},r_{h, \mathrm{th}}) < -V_\mathrm{eff}(\phi_\mathrm{fv},r_{h, \mathrm{th}})$
with $-V_\mathrm{eff}(\phi_\mathrm{tv},r_{h, \mathrm{th}})$ being no longer a local potential maximum. 
Thus around the horizon, we would never have the bounce solution that reaches the false vacuum. 
However, the thermal fluctuation becomes weaker at a distant region, and the values of $-V_{\rm eff}$ near the true vacuum become positive so that it eventually reaches $\phi = \phi_{\rm fv}$ at infinity. The schematic picture of the evolution of the scalar field is shown in Fig. \ref{flipped potential}.

To find such a solution, we numerically calculate bounce equations (\ref{bounce eq3}) and (\ref{bounce eq4}) in the shooting method. 
The calculation starts from  the horizon $r=r_{h,\mathrm{th}}$ with the boundary conditions, 
\begin{align}
\mu(r_{h,\mathrm{th}})=\mu_{\mathrm{th}-}= r_{\mathrm{th}-}/2, \quad \phi_0 \equiv \phi(r_{h,\mathrm{th}}),
\end{align}
which determines the boundary condition for the first derivative of the scalar field at the horizon
\begin{align}\
\phi'(r_{h,\mathrm{th}})=\frac{r_{h,\mathrm{th}}\frac{\partial V_\mathrm{eff}}{\partial\phi}(\phi_0)}{1-8\pi {r_{h,\mathrm{th}}}^2 V_\mathrm{eff}(\phi_0)}. 
\end{align}
We search $\phi_0$ such that $\phi$ satisfies the condition Eq.~(\ref{boundary}) at infinity
by using the shooting method implemented in Ref.~\cite{Burda:2016mou} after performing the same coordinate transformation as Eq.~\eqref{tortoise} to improve the numerical behavior near the BH horizon. Note that the bounce solution obtained from Eqs. (\ref{bounce eq3}) and (\ref{bounce eq4}) is static, the bounce action $B_\mathrm{th}$ 
is evaluated by the change of Bekenstein-Hawking entropy~\cite{Burda:2015yfa,Burda:2016mou} 
\begin{equation}
B_\mathrm{th}=4 \pi (M_+^2-\mu_{\mathrm{th}-}^2)\label{bounce thermal}.
\end{equation}

\begin{figure}
\centering
\includegraphics[width=5.6cm]{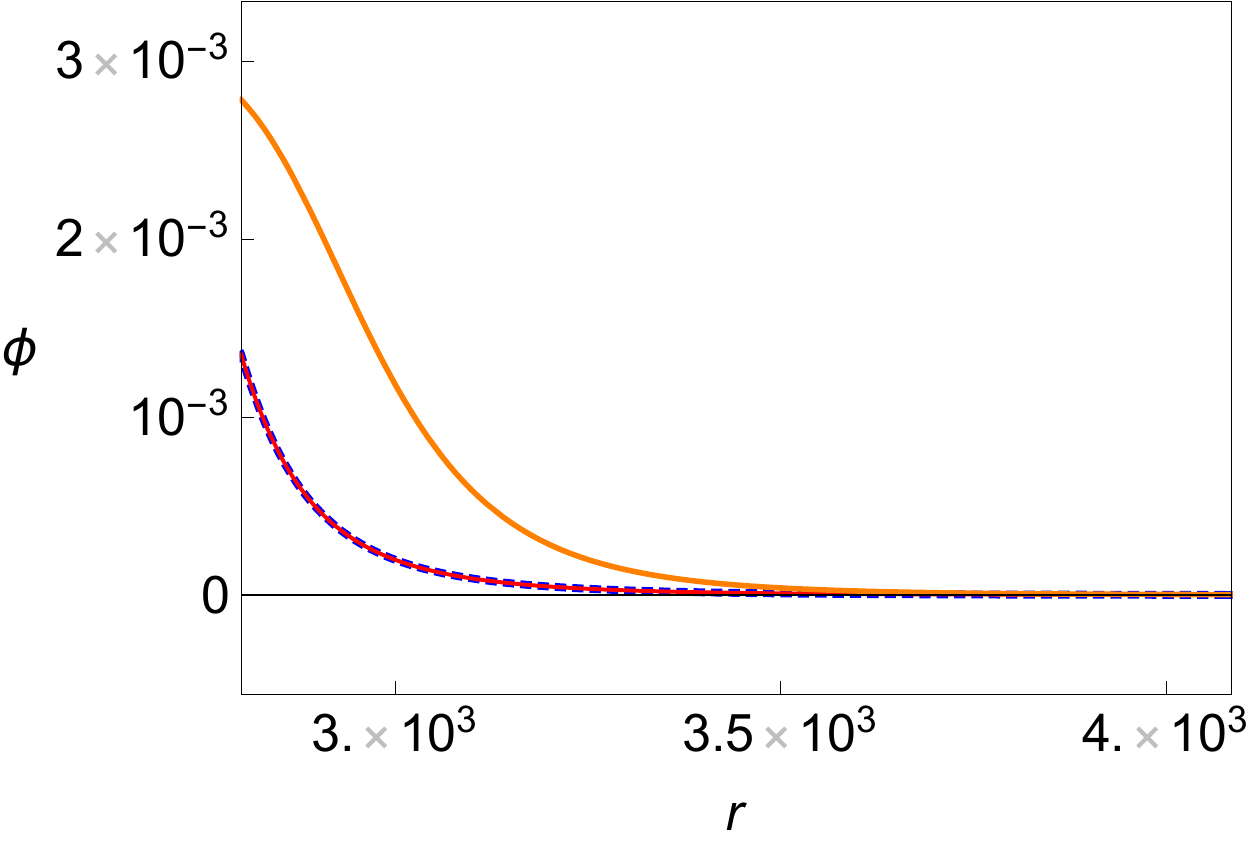}
\includegraphics[width=9.4cm]{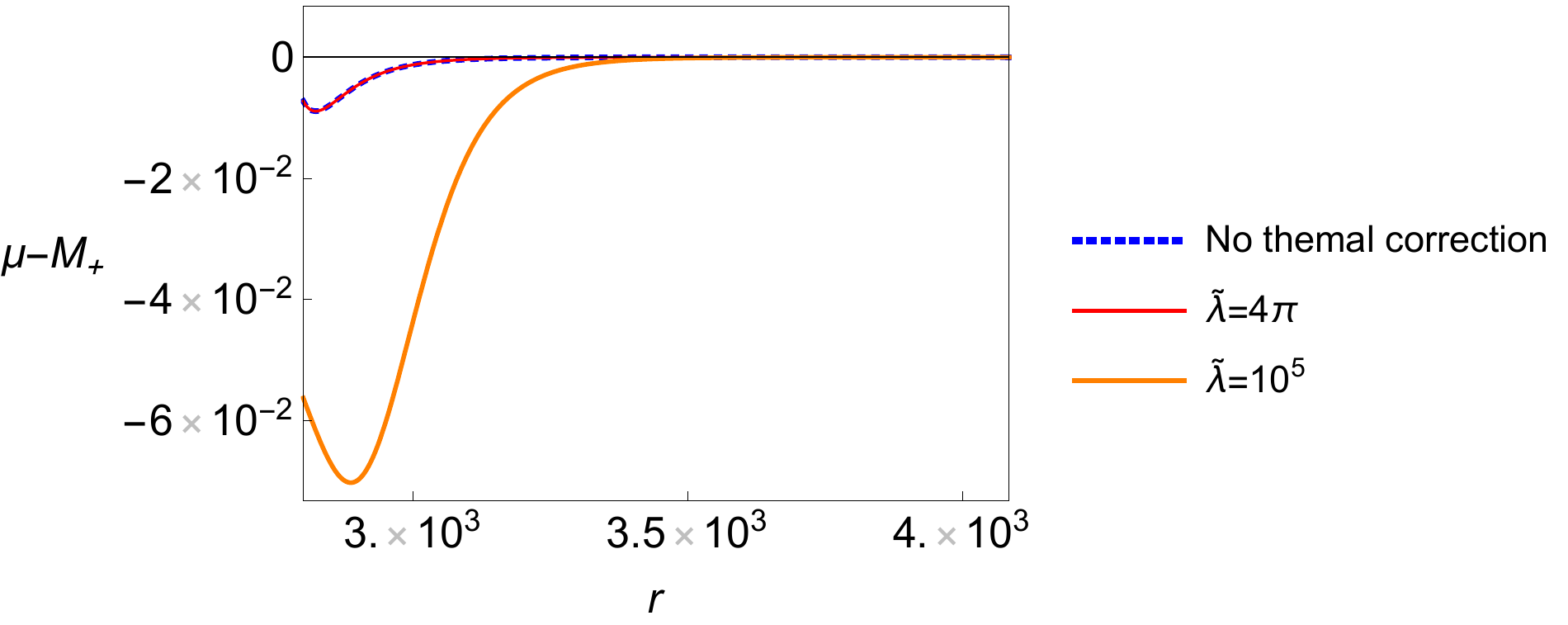}
\caption{The bounce solutions in the case with $\lambda=4\pi$, $m=2 \times 10^{-3}$, $g=3$ (or $\Delta g = 0.88$) for the toy potential $V(\phi)$, and $M_+=1.3\times10^3$. 
The blue dotted line represents the bounce solution without the thermal correction. The red and orange solid line represent those with the thermal correction for $\tilde \lambda=4\pi$ and $\tilde \lambda=10^5$, respectively.}
\label{bouncesolution}
\end{figure}

\begin{figure}
\centering
\includegraphics[width=15cm]{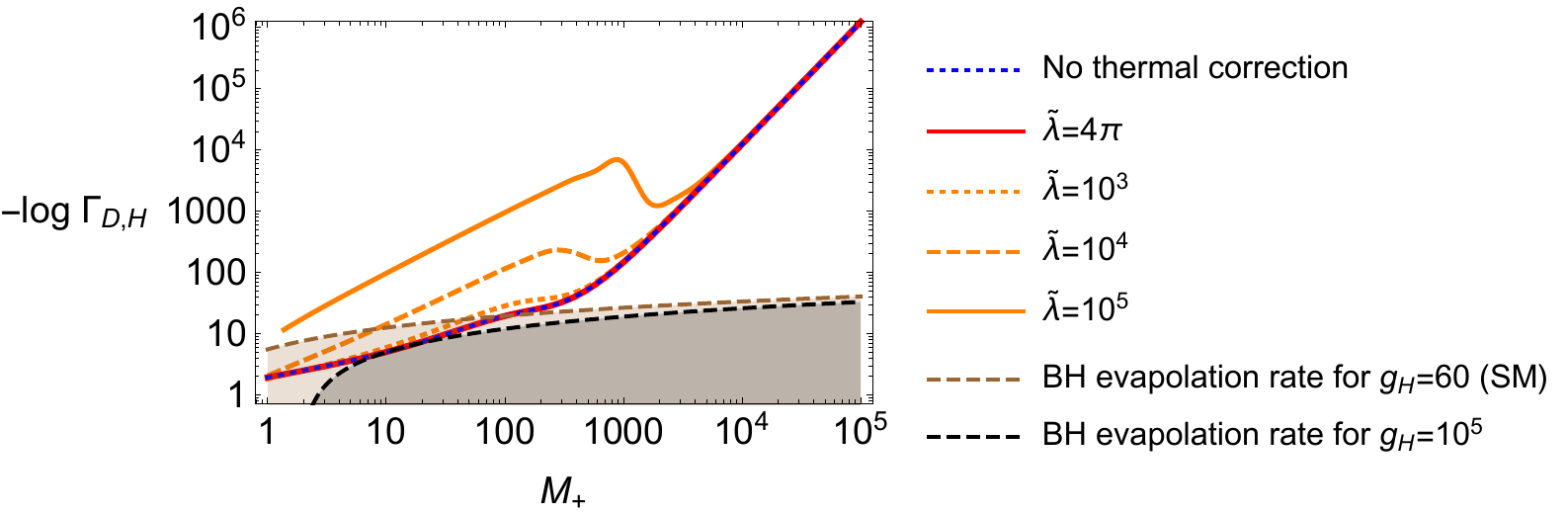}
\caption{The vacuum decay rate as a function of $M_+$ in the case with $\lambda=4\pi$, $m=2 \times 10^{-3}$, and $g=3$  (or $\Delta g = 0.88$) for the toy potential $V(\phi)$ (Eq.~\eqref{potential}) is shown. 
The blue dotted line is the decay rate without the thermal correction. The decay rates including thermal corrections with ${\tilde \lambda} = 4\pi$ (red solid), $10^3$ (orange dotted), $10^4$ (orange dashed), and $10^{5}$ (orange solid) are plotted. The orange lines represent the case where there are many particles interacting with the scalar field. For comparison, the evaporation rates with $g_H = 60$ (brown dashed) and $g_H =10^5$ (black dashed) are also shown. The former one corresponds to the case of the SM. If a line of $\Gamma_D$ comes below that of $\Gamma_H$ (shaded regions), a phase transition would occur before the BH evaporates.
}
\label{manyparticle}
\end{figure}

Performing the numerical calculations, we find that the thermal effect on the bounce solution is less significant for $\tilde{\lambda} \ll 10^3$. However, if we take ${\tilde \lambda} \gtrsim 10^3$,  
the bounce solution and action become significantly different from the one without thermal potential, 
and the decay rate is sufficiently reduced. For example, in the case $m=2\times 10^{-3}$, $g=3$, $\lambda=4\pi$, and $M_+ =1.4 \times 10^3$, we do not see the difference in the bounce solutions  for ${\tilde \lambda} = 4 \pi$ whereas we see the significant changes for ${\tilde \lambda} = 10^5$, as can be seen in Fig.~\ref{bouncesolution}. Fig.~\ref{manyparticle} shows the vacuum decay rate $\Gamma_D \equiv \sqrt{B_\mathrm{th}/2\pi} M_+^{-1} \exp[-B_\mathrm{th}]$
for $m=2\times 10^{-3}$, $g=3$, $\lambda=4\pi$ as a function of $M_+$. The condition to ignore the effect of AdS curvature on the vacuum polarization, $M_+\lesssim l=1.4\times10^5$, is satisfied. Note that if the $\Gamma_D$ is larger than the BH evaporation rate $\Gamma_H$ (Eq.~\eqref{evaporation rate}), 
the unwanted vacuum decay occurs before the BH evaporation and we would suffer from the catastrophe. 
We can see that without thermal corrections, the catastrophic vacuum decay for $M_+<10^2$ would be inevitable and the situation does not change for ${\tilde \lambda} < 10^4$. If we are allowed to take extremely large ${\tilde \lambda}>10^5$, the thermal correction can stabilize the false vacuum state so that $\Gamma_D<\Gamma_H$ is satisfied even at $M_+=1$.

\subsection{Thin-wall approximation of the bounce around  a radiating BH}
The numerical calculation performed in the previous subsection implies that  thermal effect on the bounce solution is not significant for ${\tilde \lambda} \lesssim 10^3$, regardless of the details of the other potential parameters and BH mass. 
In this subsection, in order to support this conjecture, we give an analytic investigation with the thin-wall approximation. In the thin-wall approximation, thermal correction to the bounce solution can be represented by corrections to the wall tension
$\sigma$, the AdS radius $l$, and the bubble radius $R$. It is also convenient to see how the bounce equation 
changes due to the thermal correction, as we will see later. 
%For the fixed $M_+$, in order to distinguish the properties before and after taking into account the thermal effect, we denote the latter with the subscript ``th''.
In the following, the subscript ``th'' denotes the quantities including the thermal correction.

Let us consider the case for $\overline \sigma l \ll 1$. In the case of $M_+ \ll 1/(2 \alpha)$ without the thermal correction, for which the BH catalysis effect is significant, the bubble radius is given as $R_*\simeq 2/(3 \alpha_+) -M_+/2 \simeq (2/3) \overline \sigma l^2$. 
The thermal correction to the potential at the bubble wall is $V_\mathrm{th}(\phi) = {\tilde \lambda} \phi^2/768 \pi^2 {\overline \sigma}^2 l^4$. 
If it is much smaller than the zero-temperature potential, $\Delta V$, which holds when ${\tilde \lambda}$ is sufficiently small, 
the bounce configuration as well as the wall tension do not change significantly.
%in the presence of thermal correction, when we suppose the same BH mass before the bubble nucleation, $M_+$.
On the other hand, the remnant BH mass
%after the bubble nucleation inside the bubble 
slightly changes as well as the bounce action. 
This contribution can be estimated by examining the change of the bounce equation for $\mu$, see Eqs.~(\ref{bounce eq2}) and (\ref{bounce eq4}). 
With the same field configuration for $\phi$ ($\phi = \phi_\mathrm{tv} (r_h<r<R)$ and $\phi = \phi_\mathrm{fv} (r>R)$), we obtain 
\begin{align}
&\mu_\mathrm{th}'-\mu'=\frac{{\tilde \lambda}}{192\pi }\phi^2.
\end{align}
%where $\mu$ and $\mu_\mathrm{th}$ indicate that they are evaluated with and without the thermal potential, respectively. 
Integrating this equation from horizon to infinity, 
%and regarding bubble as step function with a jump at $r=R(\gg r_h)$, 
we obtain the horizon radius of the remnant BH and bounce action with the thermal correction as
\begin{align} 
&\mu_{\mathrm{th}-}\sim  \mu_-+\frac{\tilde \lambda}{192 \pi} \phi_\mathrm{tv}^2 R \simeq \mu_-+\frac{{\tilde \lambda} m}{16\sqrt 2\pi \lambda \Delta g},\label{Bthermal1_2}\\
&B_{\mathrm{th}}\ \sim B+ \frac{ {\mu_-\tilde \lambda}  }{24}\phi_\mathrm{tv}^2 R \simeq B\left(1+\frac{3 {\tilde \lambda} \Delta g}{64\sqrt2\pi^2}\right),\label{Bthermal1}
\end{align} 
where we have used the explicit form of $\phi_{\rm tv}$ and Eq. (\ref{radiuslarge}) to obtain the right hand side of Eqs. (\ref{Bthermal1_2}) and (\ref{Bthermal1}). The condition that the thermal correction of the potential at the bubble wall is less significant can be expressed as
\begin{equation}
{\tilde \lambda} \ll \frac{128 \sqrt{2} \pi^2}{\Delta g} \simeq 2 \times 10^4 \left( \frac{0.1}{\Delta g} \right). \label{lambdacond1}
\end{equation}
We can see that as long as the constraint Eq.~\eqref{lambdacond1} is satisfied, the correction on the bounce action Eq.~\eqref{Bthermal1} 
is at most order of the unity. 
This is consistent with Fig.~\ref{manyparticle} where the significant change in the bounce action is seen only for ${\tilde \lambda}\gg 10^3$, 
although the thin-wall approximation does not hold for the parameters in Fig.~\ref{manyparticle}. 
In particular, for ${\tilde \lambda} \lesssim 4 \pi$, we find that the thermal effect cannot be significant to change the vacuum decay rate.

\par For $M_+\gg 1 /(2\alpha)$ the bubble wall nucleates near the horizon when we do not take into account the thermal correction. 
The thermal effect at the true vacuum near the horizon is $V_\mathrm{th} (\phi_\mathrm{tv}) \simeq {\tilde \lambda} \phi_\mathrm{tv}^2/3072\pi^2  M_+^2$ and is small, once more, when the ${\tilde \lambda}$ is sufficiently small,  ${\tilde \lambda} \ll 3072 \pi^2 M_+^2 \Delta V/\phi_\mathrm{tv}^2$. 
In this case the bounce configuration does not change significantly. 
Since the thermal correction is important only near the horizon and it does not affect the wall configuration, 
the slight change of the system in the thin-wall approximation can be absorbed in the change of the true vacuum energy density, $\delta V = V_\mathrm{th} (\phi) \simeq {\tilde \lambda} \phi^2/3072\pi^2  M_+^2$. 
Therefore, the thermal correction mainly changes the AdS radius, $l \rightarrow l+\delta l$, 
with 
\begin{equation}
    \frac{\delta l}{l} = \dfrac{{\tilde \lambda} \phi_\mathrm{tv}^2/3072\pi^2 M_+^2}{2 \Delta V},
\end{equation}
which changes the bounce action $B \rightarrow B+\delta B$ with $\delta B/B = 2 \delta l/l$ (see Eq.~\eqref{B1}, which reads $B \propto l^2$).
With the toy potential Eq.~\eqref{potential}, it is expressed as 
%
%Since horizon and bubble radius are very close to $2M_-$, the thermal collection of the potential can be considered as constant and absorbed by redefinition of potential parameters;
%\begin{align} 
%V_\mathrm{eff}(\phi;r)&\sim A\left(\left(\frac1 2+\frac{1}{128\pi^2 R_h^2 v^2}\right)\left(\frac{\phi}{v}\right)^2  -\frac g 3\left(\frac{\phi}{v}\right)^3+\frac 1 4 \left(\frac{\phi}{v}\right)^4\right)\\
%&=:A'\left(\frac{1}{2} \left(\frac{\phi}{v'}\right)^2-\frac{g'}{3}\left(\frac{\phi}{v'}\right)^3+\frac{1}{4}\left(\frac{\phi}{v'}\right)^4  \right)
%\end{align}
%where
%\begin{align} 
%v'=v\left(1+\epsilon\right)^{\frac1 2},\  g'=g\left(1+\epsilon\right)^{-\frac1 2},\  A'=A\left(1+\epsilon\right)^2 \quad \left(\epsilon=1/256\pi^2M_+^2v^2\ll 1\right),
%\end{align}
%then the bounce action with Hawking radiation is obtained for $g\sim g_0$:
\begin{align} 
B_{\mathrm{th}}\ \sim B\left[1+\frac{{\tilde \lambda} \Delta g}{288\sqrt2\pi^2}\left(\frac{1/2\alpha}{M_+}\right)^2\right].  \label{Bthermal2}
\end{align}
Note that the condition that the bounce configuration does not change much is given by
\begin{align}
&{\tilde \lambda} \ll 288 \sqrt{2}\pi^2 \Delta g^{-1} \left(\frac{M_+}{1/2 \alpha} \right)^2 \simeq 4 \times 10^4 \left( \frac{0.1}{\Delta g} \right) \left( 2 \alpha M_+ \right)^2.
    \label{lambdacond2}
\end{align}
This is also consistent with Fig.~\ref{manyparticle}, which shows that the deviation of the bounce actions $B$ and $B_\mathrm{th}$ starts at $M_+/(1/2\alpha) \simeq 10^{1/2}$ and 10 for ${\tilde \lambda} = 10^4$ and $10^5$, respectively.

\begin{figure}
\hspace{1mm}
\adjustbox{valign=T}{\subfigure{(a)}}
\adjincludegraphics[height=4.3cm,valign=T]{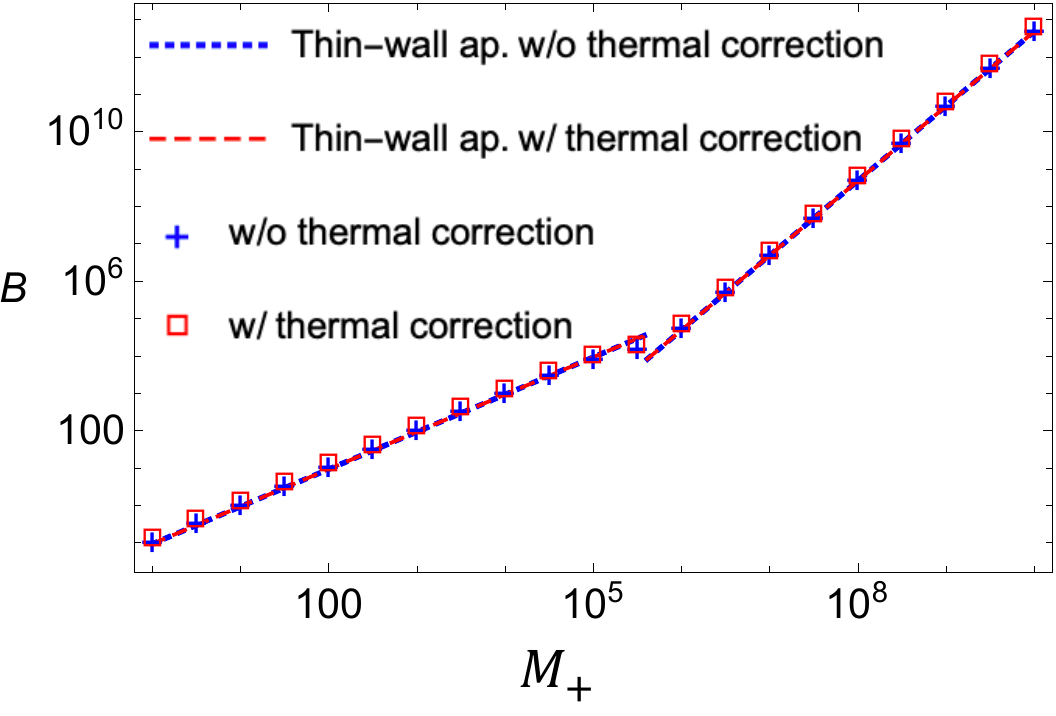}
\hspace{1mm}
\adjustbox{valign=T}{\subfigure{(b)}}
\adjincludegraphics[height=4.3cm,valign=T]{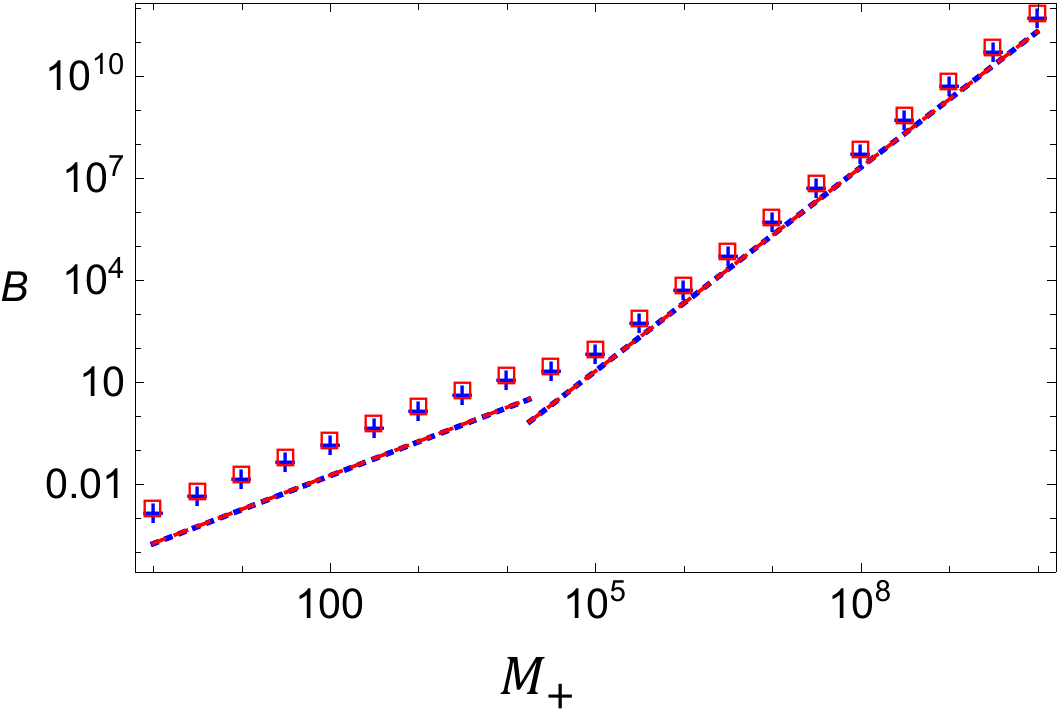}\\
\hspace{1mm}
\adjustbox{valign=T}{\subfigure{(c)}}
\adjincludegraphics[height=4.3cm,valign=T]{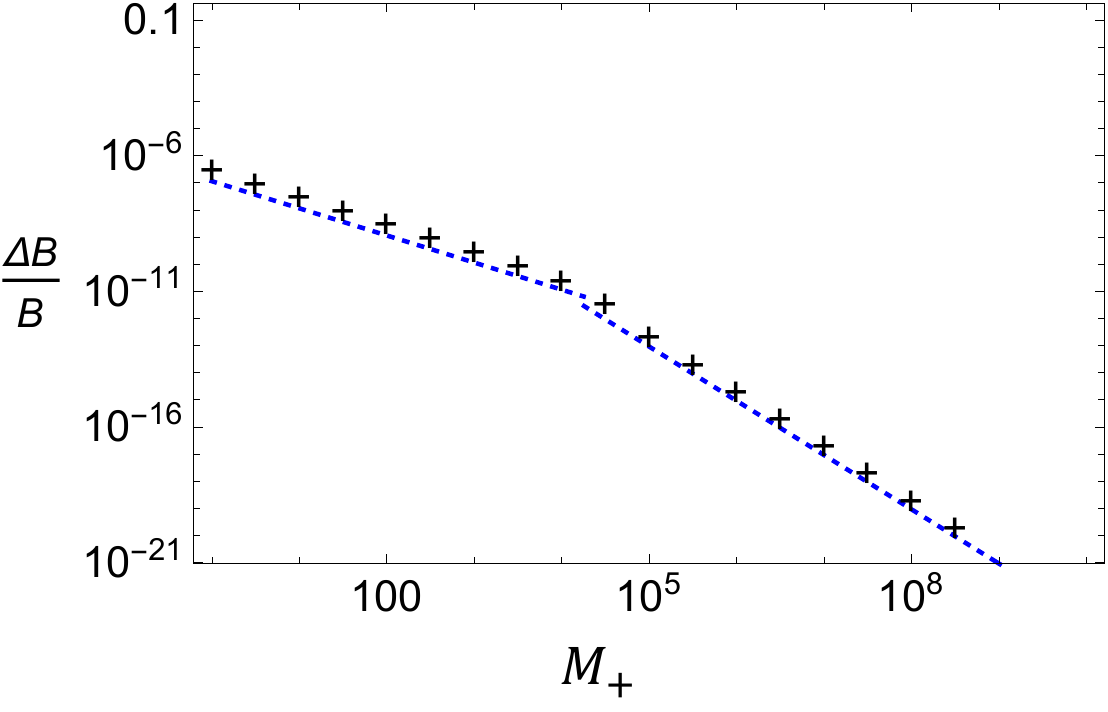}
\hspace{1mm}
\adjustbox{valign=T}{\subfigure{(d)}}
\adjincludegraphics[height=4.3cm,valign=T]{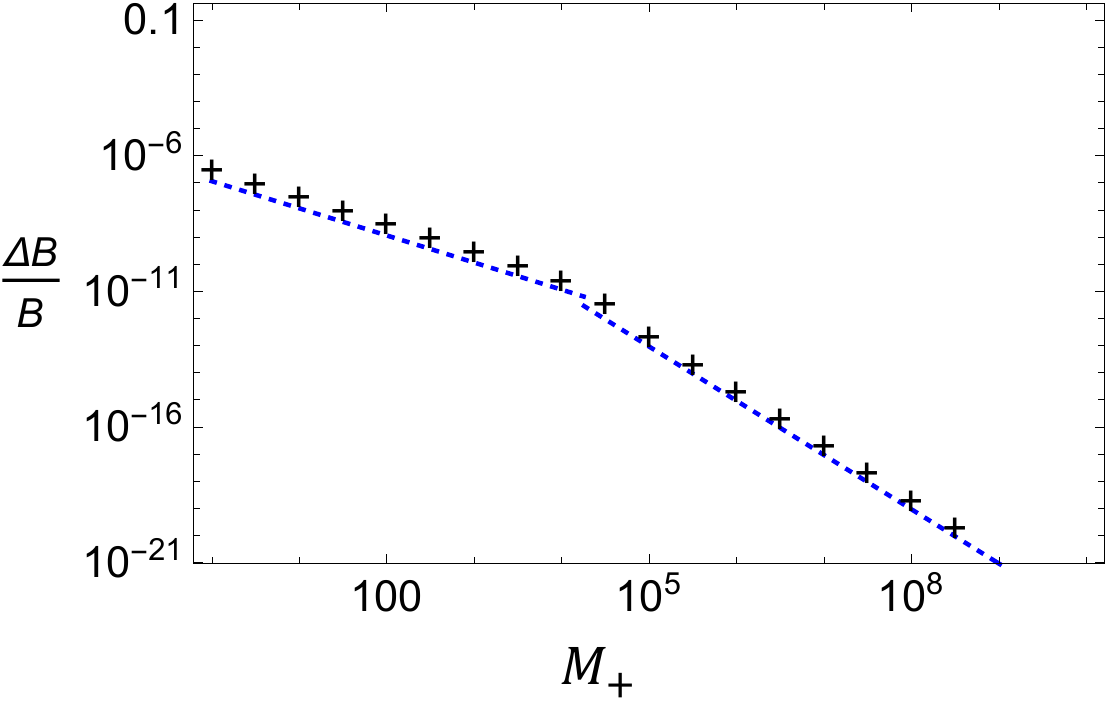}
\caption{Figures (a) and (b) show the bounce actions as functions of seed BH mass with the thin-wall ($\Delta g = 0.04$) and thick-wall ($\Delta g = 0.88$) cases, respectively. We set $\lambda={\tilde \lambda} = 4\pi$ and $m=1.45 \times 10^{-5}$. The approximate analytic formulae of the Euclidean action with/without the thermal correction (red dashed/blue dotted) are also plotted in (a) and (b). For the explicit forms of the analytic formulae, see Eqs.~(\ref{BofM0}), (\ref{Bthermal1}) and (\ref{Bthermal2}). The values of numerically obtained Euclidean actions with and without the thermal correction are plotted with the blue crosses and red squares, respectively.
The differences of the bounce actions with and without the thermal correction are shown in (c) and (d) with the black crosses, along with the thin-wall approximation formulae (Eqs.~(\ref{Bthermal1}) and (\ref{Bthermal2})) with the blue dotted lines.
}
\raggedright
\label{BofM}
\end{figure}

We find the analytic expressions that work well even when the thin-wall approximation is not valid. 
We performed a numerical calculation based on the shooting method in order to find a thick-wall bounce solution.
Figure~\ref{BofM} shows  the bounce action
with and without the thermal correction 
as a function of $M_+$. We set $\lambda = \tilde{\lambda} = 4 \pi$ and $m = 1.45 \times 10^{-5}$, and perform the numerical calculation for the thin-wall ($\Delta g = 0.04$) and thick-wall ($\Delta g = 0.88$) cases.
%for ${\tilde \lambda}=4 \pi$ and for the toy potential (Eq.~\eqref{potential}) with $\lambda = 4\pi$ and $m=1.45 \times 10^{-5}$.
For comparison, we also plot the analytic formulae of the Euclidean action, derived in the thin-wall approximation, in Figure~\ref{BofM}-(a) and (b) (see Eqs.~(\ref{BofM0}), (\ref{Bthermal1}), and (\ref{Bthermal2})). The analytic formulae are consistent with the numerical results and the thermal correction do not give significant differences in the bounce actions
for $\lambda = {\tilde \lambda} = 4 \pi$.
The increments of bounce actions due to the thermal correction, $\Delta B$, are too small to see in Figure~\ref{BofM}-(a) and (b).
%they are actually explained by the analytical expressions as shown in Figure~\ref{BofM}-(c) and (d).
Figure~\ref{BofM}-(c) and (d) plot $\Delta B/B$ for the thin-wall and thick-wall cases, respectively, and both are well consistent with the formulae derived by the thin-wall analysis presented in Eqs.~(\ref{Bthermal1}) and (\ref{Bthermal2}).
Therefore, we conclude that the thermal correction cannot rescue the Universe from the unwanted vacuum decay catalyzed by the BH 
unless the effective coupling to the Hawking particles, ${\tilde \lambda}$, is extremely large (Eqs.~\eqref{lambdacond1} and~\eqref{lambdacond2}).

\section{Implication to the Higgs instability} \label{sec4}
As an application of our findings in the previous section, 
we investigate the SM Higgs vacuum instability catalyzed by a BH. 
It has been studied in Ref.~\cite{Burda:2016mou} without thermal correction, 
which suggests that we suffer from the vacuum decay from the electroweak vacuum to the AdS true vacuum, 
which is catalyzed by a BH with a mass $M_+ \lesssim 10^7$ if the Standard Model Higgs potential 
is negative at the field values larger than $10^{11}$ GeV. 
We here take into account the thermal correction to see if it can rescue our Universe from this catastrophe.

The SM Higgs potential at field values much larger than the electroweak scale is well described by 
\begin{align}
V_\mathrm{eff}^h (\phi)=\frac {\lambda_h(\phi)} 4 \phi^4\label{VHiggs}
\end{align}
in the Unitary gauge.
Here $\phi$ denotes the SM Higgs field and $\lambda(\mu)$ is the running Higgs quartic coupling. 
Instead of solving the renormalization group equations for $\lambda_h$ from the electroweak to high energy scales~\cite{Degrassi:2012ry,Buttazzo:2013uya}, 
we here adopt the fitting formula used in Ref.~\cite{Burda:2016mou}
\begin{align}
\lambda_h(\mu)=\lambda^*+b\left(\log\left(\frac \mu {M_\mathrm{pl}}\right)\right)^2+c\left(\log\left(\frac \mu{M_\mathrm{pl}}\right)\right)^4. \label{higgsV}
\end{align}
Here
we adopt $\lambda^*=-0.013, b=1.3\times 10^{-5}, c=1.7\times 10^{-6}$ which models the Higgs potential with the running coupling for top mass $m_t=173\mathrm{GeV}$ and Higgs mass $m_h=125\mathrm{GeV}$. 
For these parameters, the Higgs potential becomes negative at $\phi = \phi_c \sim 10^{10}$ GeV $\sim 10^{-9}$ in the Planck unit.

We now study whether the potential correction from thermal fluctuation can stabilize the false vacuum enough to prevent the vacuum phase transition. 
It is complicated to calculate a 1-loop effective potential of a scalar field in the Schwarzschild background when it interacts with fermions or gauge fields. 
Thus we instead approximate the correction by the thermal mass for the SM Higgs~\cite{Giudice:2003jh,Buttazzo:2013uya}, 
\begin{align} 
m_T^2&\sim \left(\frac 3 {16} g_2^2+\frac 1 {16}g_Y^2+\frac 1 4 y_t^2 +\frac 1 2 \lambda_h \right)T^2 \label{higgsT}
%&\sim0.092 T^2,
\end{align} 
with $T$ being replaced by the ``$r$-dependent Hawking temperature'', $T_H = 1/4  \pi r$, see Eq.~\eqref{thermalpot}. 
Here $g_2,\ g_Y,$ and $y_t$  are the $SU(2)_L,\ U(1)_Y$ gauge couplings, and the top yukawa coupling, respectively. 
Note that the contributions from other particles are negligibly small. 
Let us choose the renormalization scale for the coupling constants as the Planck scale since we are interested in the phase transition induced by a tiny BH whose Hawking temperature is close to the Planck scale. 
Then we get $(3/16)g_2^2+(1/16) g_Y^2+(1/4) y_t^2 +(1/2) \lambda_h  \simeq 0.092$ and the effective potential of Higgs field around a BH is estimated as 
\begin{align} 
V_\mathrm{eff}^h (\phi;r)\sim V_\mathrm{eff}^h (\phi)+  \frac {0.092} {32\pi^2 r^2}  \phi^2,  \label{higgs thermal potential}
\end{align}
which corresponds to the effective coupling ${\tilde \lambda} \simeq 2.2$ for which the thermal correction is negligible.

\begin{figure}
\centering
\includegraphics[width=16cm]{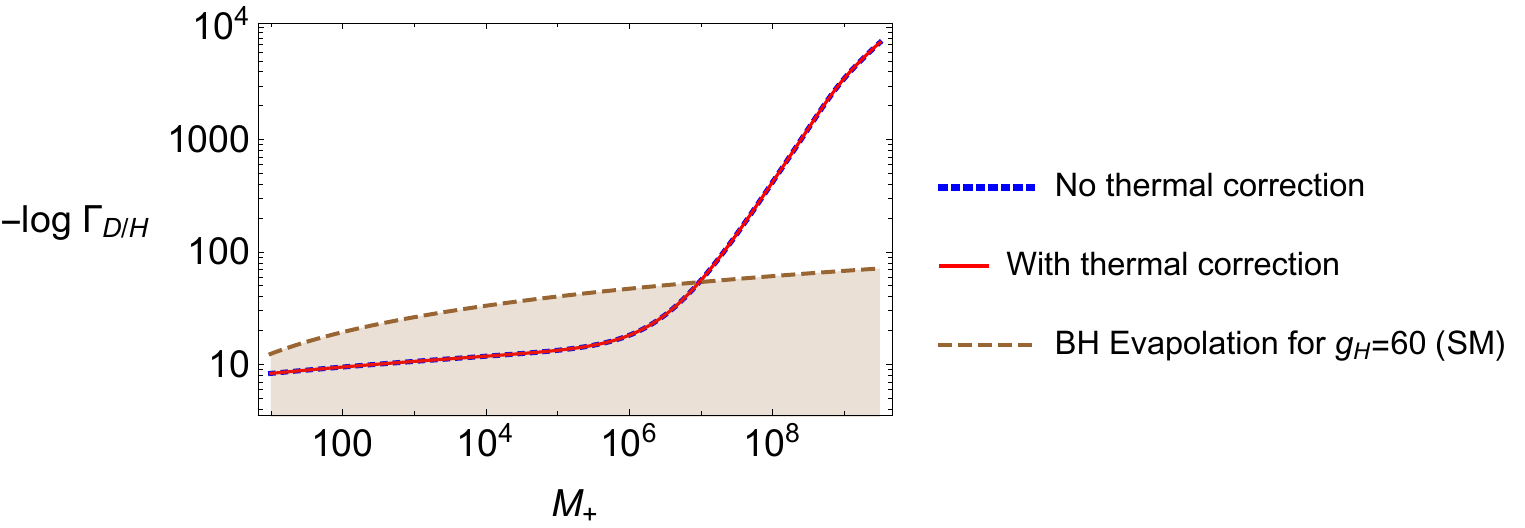}
\caption{The bubble nucleation rate for the SM Higgs field with a BH as a function of the seed BH mass $M_+$ is shown.  Here 
we take the potential parameters as $\lambda^*=-0.013, b=1.3\times 10^{-5}, c=1.7\times 10^{-6}$. The blue dotted line indicates the bubble nucleation rate without thermal correction, while red solid line indicates the one with the thermal correction (Eq.~\eqref{higgs thermal potential}). Brown dashed line represents the BH evaporation rate. 
The bubble nucleation rate exceeds the BH evaporation rate in the brown region, which means that the Universe would undergo 
the Higgs vacuum decay for $M_+<10^7$.}
\label{BofMHiggs}
\end{figure}

With this effective potential, we solve the bounce equations in the same way as discussed in Sec.~\ref{sec3}, see Eqs.~\eqref{bounce eq3} and~\eqref{bounce eq4}. 
Figure~\ref{BofMHiggs} shows the numerical results for the bounce action. 
We can see that the thermal correction does not change the phase transition rate practically and the bubble nucleation rate is larger than the BH evaporation rate for smaller BH mass with $M_+ \lesssim 10^7$, which means that the Higgs vacuum decay would be inevitable if microscopic BHs exist in the Universe\footnote{One way around would be to lower the Planck scale by the order of $10^{-7}$ so that the phase transition rate is well suppressed at the energy scales where the semi-classical approximation is valid.}.
Note that the bounce action is proportional to the seed BH mass $M_+$~\cite{Burda:2016mou}, similar to the case in the thin-wall approximation for the small $M_+$~(Eq.~\eqref{smallmbounceaction}). 
At $M_+ \simeq 10^7$, the bounce action is of the order of unity and we do not have the exponential suppression in the vacuum decay rate for smaller BH mass.

This behavior can be understood in a similar way to the previous sections. 
In order for the thermal effect to be significant enough to change the bounce action, 
the thermal correction at a bubble wall should be comparable to or larger than the zero-temperature potential energy density.
Since the zero-temperature potential is roughly described by a scale invariant potential, 
$V =- |\lambda_h|\phi^4/4$
with $\lambda_h \sim -0.01$, the $O(4)$ bounce solution are found to be scale invariant~\cite{Arnold:1991cv},
and the scale of the weak violation of the scale invariance determines the minimum bounce solution that gives the minimum bounce action. 
Since now we are working on the $O(3)$ bounce solution with a typical scale $\phi_c$,
we expect that the field value inside the bubble that gives the dominant contribution for the vacuum decay 
is roughly $\phi_0 \sim \phi_c$. 
Then from the balance between the bulk and wall bounce action, 
the bubble radius is roughly estimated as $R \sim 1/\sqrt{|{\lambda_h}|}\phi_c$ (see Eq.~\eqref{estimateR}). As a result, the thermal effect of the potential at the wall is estimated as ${\tilde \lambda} |\lambda_h| \phi_c^4 /768 \pi^2$, 
whereas the negative vacuum energy inside the bubble is estimated as $- |\lambda_h| \phi_c^4/4$. 
Thus the thermal effect is strong enough to change the bounce configuration if ${\tilde \lambda} \gg 192 \pi^2 \simeq 2 \times 10^3$.
The Higgs potential, leading to the bubble configuration,
%can be parameterized by the height and width of the potential barrier and the potential energy at $\phi=\phi_0$ and
can be modeled by our toy model with $\lambda\sim |\lambda_h|$, $\phi_\mathrm{tv}\sim\phi_c$, and $\Delta g\sim O(1)$. Actually, the condition for having the strong thermal effect is consistent with that derived in our toy model shown in Eq.~\eqref{lambdacond1}.
Note that in this configuration, the AdS radius is given as $l\sim\sqrt{1/ |\lambda_h|\phi_c^4}$ 
whereas the wall tension is estimated as $\overline{\sigma} \sim  |\lambda_h|^{1/2} \phi_c^3$.
From Eq.~\eqref{smallmbounceaction}, we evaluate the bounce action as $B \sim \phi_c M_+/|\lambda_h|^{1/2}$, 
which is consistent with the numerical result (also see Ref.~~\cite{Burda:2016mou}). 
\par One might consider that the thermal correction can rescue the Universe from the unwanted vacuum decay 
if there are sufficiently large number of hidden fields that couple to the Higgs field to give a large effective coupling ${\tilde \lambda}$. 
However, such fields easily change the renormalization group equation running for the Higgs quartic coupling $\lambda_h(\mu)$, 
whose effect is much more significant than the thermal effect we consider in this work. 
Therefore we conclude that the BH catalysis effect on the vacuum decay is inevitable for the SM Higgs potential. 
Our very existence suggests that even a tiny primordial BH with the mass smaller than $\sim 10^7$ had never created in the observable Universe.
As other possibilities, the SM Higgs potential may be stabilized by a new physics or is just stable with a relatively light top mass~\cite{Aaboud:2018zbu,Khachatryan:2015hba,Bezrukov:2014ina}.

\section{Conclusions and Discussion} \label{sec5}
It has been discussed that a small BH, whose mass is smaller than $M_+ < 10^7$, catalyzes the bubble nucleation in a false vacuum of the Higgs field and its decay rate is evaluated by the bounce action with the zero-temperature potential~\cite{Gregory:2013hja,Burda:2015yfa,Burda:2016mou}
and that the rate can be significantly higher than the CdL decay rate. However, such a small BH has its high Hawking temperature, $T_{\rm H} \gtrsim 10^{12}$ GeV, for which the effective potential of true vacuum is lifted to prevent the system from the vacuum decay.

In this work, we have evaluated 
a static $O(3)$ bounce action around a Schwarzschild BH in a false vacuum state. Using the Unruh vacuum state, we have taken into account a thermal mass which has its radial-dependence and is proportional to $r^{-2}$ 
with $r$ being the distance from the BH. 
The numerical results show that the Euclidean action increases due to the thermal correction, 
but it is negligibly small effect for ${\tilde \lambda} \lesssim 4 \pi$. In our setup, the thermal effect can be significant when there exist a number of scalar fields couple to the metastable field and ${\tilde \lambda} \gg 10^3$ is satisfied. 
This is because even though the Hawking temperature can be high near the horizon,
it is suppressed at the bubble wall 
and hence it cannot change the bounce configuration much. 
With the help of the thin-wall approximation, we have derived the formula that describes the small increase of the bounce action, 
which remarkably works well even for the case where the thin-wall approximation does not hold. 
This also gives the condition that the thermal correction can be important. That is, in order for the thermal correction to be significant, the thermal correction should be larger than the zero-temperature potential at least
at the true vacuum around the bubble wall so that the bounce configuration is significantly changed. 
This requires an extremely large effective coupling between the scalar field and the Hawking radiation, 
${\tilde \lambda} \gg 10^4$, since the two-point function $\bra{U} \phi^2 \ket{U}$ is suppressed by a small factor $(192\pi^2)^{-1}$(see Eq.~\eqref{vacuum polarization}). 
Note that we have assumed that the quadratic thermal mass term holds even at the true vacuum around the bubble wall. However, if the effective Hawking temperature is lower than the expectation value of the true vacuum, the thermal fluctuation is negligible as discussed in the context of the Affleck-Dine mechanism for baryogenesis~\cite{Anisimov:2000wx,Fujii:2001zr,Kasuya:2001hg,Chiba:2010ff}. In such a case the thermal effect would become much smaller, yet our main conclusion is unchanged that thermal effect does not alter the bounce action significantly for a reasonable coupling between the scalar field and Hawking radiation.

We have also applied our calculation to the SM Higgs vacuum instability~\cite{Arnold:1989cb,Sher:1988mj,Arnold:1991cv,Espinosa:2007qp} around a BH~\cite{Burda:2016mou}. 
Thermal correction to the potential (Eq.~(\ref{higgs thermal potential})) is estimated by the thermal mass obtained from the SM particle contents with the temperature being replaced by 
the $r$-dependent effective Hawking temperature. We have found out that the effective coupling between the SM Higgs and the Hawking 
radiation can be modeled by ${\tilde \lambda} \simeq 2.2$ in our toy model and that the thermal effect on the bounce action is negligibly small. 
As a result, the phase transition rate 
exceeds the BH evaporation rate for small BH, $M_+ < 10^7$, which is the catastrophe of our Universe. 
Therefore we conclude that we should have never had such small primordial BHs in the observable Universe
unless the SM Higgs potential is stabilized by a new physics or is stable as it is. 
Note that the latter possibility is still consistent with the present uncertainty of the top mass~\cite{Aaboud:2018zbu,Khachatryan:2015hba,Bezrukov:2014ina}.

Finally we comment on some interpretation issue for the BH catalysis effect. Ref.~\cite{Mukaida:2017bgd} raised some possible interpretations for the BH catalysis effect. One of the interpretations is that the catalysis effect would be caused by thermal plasma around a BH and it can be understood as a sphaleron process in a uniform thermal bath\footnote{According to Ref.~\cite{Mukaida:2017bgd}, another scenario is that a vacuum bubble could be nucleated in the vicinity of the BH horizon with its non-zero kinetic energy to expand. This interpretation would work for a realistic situation where the Hawking plasma is localized near the horizon.}. However, our study shows that the thermal plasma is localized near the horizon and a nucleated bubble wall is not immersed by the plasma. Although it appears inconsistent with the proposed interpretation in~\cite{Mukaida:2017bgd}, our argument makes sense if the catalysis effect is caused by the attractive force of a BH as is explained below. The size of the bubble wall is determined by the balance among the attractive force from a BH, bubble tension, and interior vacuum energy since it is simply derived from the classical Einstein equation. Then the resulting bubble has a smaller size due to the attractive force from the BH, which reduces the Euclidean action that is calculated by the surface integration on the Euclidean bubble. Therefore, the BH catalysis effect may be caused by the attractive force, and it is not necessary that the vacuum bubble nucleated by the BH catalysis effect is immersed by thermal plasma.

%%%%%%%%%%%%%%%%%%%%%%%%%%%%%%%%%%%%%%%%%%%%%%%%%%
\section*{Acknowledgments}
%%%%%%%%%%%%%%%%%%%%%%%%%%%%%%%%%%%%%%%%%%%%%%%%%%

We thank Ruth Gregory, Keisuke Inomata, Ian Moss, Kyohei Mukaida, Davide Racco, Masaki Yamada, and Yusuke Yamada for useful discussions and comments.
The work of TH is supported by Program of Excellence in Photon Science. 
KK is supported by JSPS KAKENHI, Grant-in-Aid for Scientific Research JP19K03842 and Grant-in-Aid for Scientific Research on Innovative Areas 19H04610. 
NO is supported by the JSPS Overseas Research Fellowships and the Perimeter Institute for Theoretical Physics. Research at Perimeter Institute is supported in part by the Government of Canada through the Department of Innovation, Science and Economic Development Canada and by the Province of Ontario through the Ministry of Colleges and Universities.
JY is partially supported by JSPS KAKENHI Grant Nos.\ 15H02082, 20H00151, and Grant-in-Aid for Scientific Research on Innovative Areas 20H05248.

\end{document}